\documentclass{optica-article}

\renewcommand*{\journal}[1]{%
  \IfFileExists{#1.sty}{%
    \RequirePackage{#1}
  }{%
    \ClassError{optica-article}{You've specified an unsupported journal '#1'. Was that a typo, or are you using a copy of the template without the complete set of style files?}{Was that a typo, or are you using a copy of the template without the complete set of style files?}
  }
}

\journal{opticajournal} 
\articletype{Research Article}

\usepackage{lineno}

\newcommand{\homedir}{./hardware_pca_onn}

\usepackage{amsmath}
\usepackage{amsfonts}

\newcommand{\papertitle}{
Deep Feature-specific Imaging
}

\usepackage{enumitem,kantlipsum}

\usepackage{siunitx}

\makeatletter
\@ifpackageloaded{xcolor}{
    \PassOptionsToPackage{dvipsnames}{xcolor}
}{
    \usepackage[dvipsnames]{xcolor}
}
\makeatother
\newcommand{\rnote}[1]{\textcolor{red}{#1}}
\newcommand{\bnote}[1]{\textcolor{blue}{#1}}

\def\sgn#1{\text{sgn}\left(#1\right)}

\expandafter\let\expandafter\oldequationstar\csname equation*\endcsname%
\expandafter\let\expandafter\endoldequationstar\csname endequation*\endcsname%
\renewenvironment{equation*}%
  {\linenomath\oldequationstar}{\endoldequationstar\endlinenomath}%

\newcommand{\B}{\boldsymbol}
\newcommand{\IR}{\mathbb{R}}

\usepackage{subcaption}

\newcommand{\Xalign}[1]{\begin{aligned}#1\end{aligned}}

\newcommand{\Xequa}[1]{\begin{equation}\Xalign{#1}\end{equation}}

\def\Xhide#1{}

\def\noisy#1{\tilde{#1}}
\def\AGN{\mathcal{N}}
\def\Poisson{\mathcal{P}}
\def\Poisson{\mathrm{Poisson}}

\usepackage{graphicx} 
\def\Xpolish#1#2{#2}

\def\bnote#1{\textcolor{blue}{#1}}
\def\rnote#1{\textcolor{red} {#1}}
\def\ynote#1{\textcolor{orange}{#1}}

\def\YLnote{\rnote}

\def\ynote#1{\color{orange} #1 \color{black}}
\def\AVnote{\ynote}

\usepackage[normalem]{ulem}
\def\checkednote#1{\sout{#1}}

\usepackage{tikz}
\usetikzlibrary{shapes.geometric}
\usetikzlibrary {shapes.misc}
\usepackage{caption}
\usepackage{amsmath,amsfonts}

\def\YLnote{}

\usepackage{printlen}
\usepackage{layouts}

\usepackage{float}

\usepackage{booktabs}
\usepackage{tabularx} 

\begin{document}

\title{\papertitle} 





\author{Yizhou Lu\authormark{1,*} and Andreas Velten\authormark{2,1}}
\address{\authormark{1}Department of Electrical and Computer Engineering, University of Wisconsin-Madison, Engineering Hall, 2415, 1415 Engineering Dr, Madison, WI 53706, USA\\
\authormark{2}Department of Biostatistics and Medical Informatics, University of Wisconsin-Madison, Warf Office Bldg, 610 Walnut St Room 201, Madison, WI 53726, USA\\}
\email{\authormark{*}ylu289@wisc.edu}

\begin{abstract*}
\YLnote{
Modern photon-counting sensors are increasingly dominated by Poisson noise, yet
conventional feature-specific imaging (FSI), based on principal component analysis (PCA),
is optimized for additive Gaussian noise and variance preservation rather than task-specific
objectives, leading to suboptimal performance and a loss of its advantages under Poisson noise.
To address this, we introduce DeepFSI, what we believe to be a novel end-to-end optical-electronic
framework. DeepFSI "unfreezes" PCA-derived masks, enabling a deep neural network to learn
globally optimal measurement masks by computing gradients directly under realistic Poisson
and additive noise conditions. Simulations and hardware experiments demonstrate that DeepFSI
achieves improved classification accuracy and stronger transfer robustness compared to PCAbased FSI across varying photon budgets, particularly in Poisson-noise-dominant environments.
DeepFSI also exhibits enhanced robustness to design choices and performs well under additive
Gaussian noise, representing a significant advance for noise-robust computational imaging in
photon-limited applications.
Simulations and hardware experiments demonstrate that DeepFSI achieves improved classification accuracy and stronger transfer robustness compared to PCA-based FSI across varying photon budgets, particularly in Poisson-noise-dominant environments.
DeepFSI also exhibits enhanced robustness to design choices and performs well under additive Gaussian noise, representing a significant advance for noise-robust computational imaging in photon-limited applications.
}
\end{abstract*}

\section{Introduction}\label{sec:1_intro}

%
%
The development of advanced sensor technologies, particularly photon-counting sensors (PCS) like Single-Photon Avalanche Diodes (SPADs), has introduced unprecedented temporal precision for photon-limited applications \cite{ingle2019high, niehorster2016multi, antolovic2017spad, altmann2018quantum}. 
%
%
PCS have been increasingly explored in computational imaging (CI) frameworks to enhance performance.
CI uses optical coding followed by computational decoding \cite{cossairt2012does}, and the coding is traditionally designed for additive Gaussian noise (AGN). 
Generally, it can be expressed by 
 \begin{equation}
    \label{eqn:NoiselessMeasurement}
    \begin{aligned}
        \B{y} &=  \B{M x} ,
    \end{aligned}
\end{equation}
where $\B{y}$ incorporates the measured photon numbers at sensor, $\B{M}$ is the sensing matrix of a given coding design, and $\B{x}$ is the vector representation of an image \cite{willet2009CSPoisson, cossairt2012does, harwit1979hadamard}.
\YLnote{
    Unless otherwise stated, we assume a fixed photon budget, defined as the total expected number of photons incident on the coding device and distributed throughout the coding sequence of $\B{M}$, across all coding designs in this work.
    }
\YLnote{Because read noise is negligible} 
on PCS, signal-dependent Poisson noise (PN) dominates in most imaging scenarios,  
undermining the performance gains typically achieved by such traditional CI methods \cite{cossairt2012does}.
%
To overcome this challenge, Cossairt et al. proposed task-specific imaging, often 
exemplified by 
feature-specific imaging \cite{neifeld2003FSI}, as a promising direction for future research \cite{cossairt2012does}.

Feature-specific Imaging (FSI) utilizes prior knowledge about the scene and task, for example in the form of principal components obtained via principal component analysis (PCA) that are applied to the image before the light is measured and digitized. 
Using these coding patterns, it is possible to extract features of interest before digitization, application of \Xpolish{Poisson noise}{PN}, and digital processing.
FSI offers benefits in reducing data burden, improving efficiency, and enhancing feature fidelity in high-AGN environments \cite{neifeld2003dual}.
%
However, conventional FSI paradigms, including Neifeld et al.'s foundational work and subsequent developments, are primarily designed and optimized under AGN \cite{neifeld2003FSI, neifeld2003dual, neifeld2014optimizing, neifeld2008adaptive}.
Its performance degrades in conditions where PN becomes the primary noise source, making it suboptimal in such cases \cite{neifeld2003dual}.
%
Indeed, early investigations into FSI have explicitly shown that its multiplexing advantages from coding disappear \YLnote{under fixed photon budget} when moving from AGN to PN, resulting in no feature-fidelity improvement over conventional, non-computational imaging methods \cite{neifeld2003FSI}.
Therefore, it is essential to develop new feature extraction and measurement optimization methods for FSI that are intrinsically robust to PN, moving beyond ad-hoc noise handling at the post-processing stage. 
Specifically, for PCS, designing PN-aware algorithms for the entire FSI pipeline is becoming imperative.

\YLnote{
In this context, we propose Deep Feature-specific Imaging (DeepFSI), an end-to-end framework designed to address these challenges.
%
Traditional pipelines adhere to an "image-then-task" paradigm where hardware is optimized to capture and reconstruct an image.
%
%
In contrast, DeepFSI underscores a "task-first" perspective. 
It integrates an optical-coding layer at the hardware front-end with a deep neural network at the software back-end, optimized directly for specific computer vision tasks.
}

\YLnote{
The core motivation behind DeepFSI is to prioritize direct, photon-efficient feature extraction 
by treating full image reconstruction as a non-essential intermediary for classification tasks.
\Xpolish{
\YLnote{ [Delete]
If images are expected, the software backend can also be adapted for image recovery by replacing the classification head with reconstruction-oriented models, such as a U-Net decoder.
}
However, we argue that the "image-then-task" paradigm is sub-optimal for discriminative.
}{
    We argue that the "image-then-task" paradigm is suboptimal for discriminative tasks.
}
In PN-limited regimes with a fixed photon budget, distributing photons across a large number of measurements 
reduces the signal-to-noise ratio (SNR) of each individual measurement \cite{willet2009CSPoisson}.
Image reconstruction is an "information-dense" task that requires a much broader range of spatial features, many of which are eventually discarded in downstream classification.
%
%
%
Therefore, we redefine fidelity not as {mean squared error (MSE)}, a metric primarily associated with pixel-level visual similarity, to a ground truth, but as the preservation of task-essential information. 
%
Classification accuracy serves as a quantitative measure of how effectively the sensing matrix captures and preserves critical features and functional information in the presence of PN.
By concentrating the available photon budget on a compact set of task-relevant features, DeepFSI avoids the SNR dilution caused by over-sampling and instead focuses on a narrow, high-SNR subspace that contains the most discriminative information, thereby maximizing functional feature fidelity.
}

\YLnote{
During the training stage, DeepFSI explicitly incorporates both AGN and PN models after the optical-coding layer.
By backpropagating gradients through this noise-aware layer, the sensing matrix is optimized for robust feature extraction.
Following the experimental setup in \cite{neifeld2014optimizing}, we use a single-pixel camera (SPC) as our primary testbed.
Our results show that DeepFSI outperforms conventional FSI under PN conditions in extracting features for classification tasks.
Beyond accuracy, DeepFSI also demonstrates greater practical robustness.
Traditional FSI methods often require precise prior knowledge of the number of features and noise levels to avoid improper compression.
In contrast, DeepFSI maintains strong performance even when the number of features is mismatched, making it more reliable in real-world scenarios where scene statistics and noise levels are unpredictable.
%
Crucially, we find that the optimal mask design is significantly affected by photon noise, resulting in masks that differ from those designed under noiseless or AGN assumptions, which challenges the common practice of designing masks and optics without accounting for PN.
}

The main contributions of the proposed DeepFSI model are as follows:
\YLnote{
\begin{enumerate}
    \item \textbf{Noise-Aware Supervised Optimization}: We establish an end-to-end framework that integrates signal-dependent noise modeling directly into the FSI pipeline. 
    Unlike traditional unsupervised methods (e.g., PCA), our approach optimizes coding patterns specifically for supervised tasks, ensuring intrinsic robustness to PN. 
    This also highlights that coding patterns optimized for AGN are suboptimal in PN-limited regimes, emphasizing the need for physics-aware design.
    \item \textbf{Task-Prioritized Photon Allocation}: DeepFSI allocates the finite photon budget to task-essential features.
    By avoiding the SNR dilution caused by high-bandwidth sampling required for full image reconstruction, the framework maximizes functional feature fidelity for discriminative tasks. 
    \item \textbf{Scalable, Robust, and Transferable Architecture}: DeepFSI is compatible with state-of-the-art backbones such as Vision Transformers, demonstrating that front-end physical coding optimization delivers performance gains beyond back-end model complexity alone. 
    Extensive hardware experiments with a photon-counting single-pixel camera show that the framework maintains high fidelity across complex datasets, tolerates mismatched noise levels, and performs reliably under suboptimal compression ratios.
\end{enumerate}
}

\section{Related Work}
\label{sec:2_related}

The studies most relevant to our research are summarized below.

\begin{itemize}
    \item \YLnote{
        \textbf{Evolution in FSI}. The development of Feature-Specific Imaging (FSI) has evolved from static image reconstruction toward task-driven paradigms. 
        Originally introduced by Neifeld et al. \cite{neifeld2003FSI}, FSI was formulated as a framework that directly measures features of interest, primarily through PCA, to maximize feature fidelity under a fixed photon budget. 
        %
        While later viewed as a variant of compressed sensing  \cite{neifeld2014optimizing},
        FSI distinguishes itself by emphasizing a data-driven, task-specific design that incorporates prior statistical knowledge of the scene \cite{neifeld2003FSI}, rather than relying on generic sparsity assumptions and random projections.
        A key development of FSI was the Adaptive FSI (AFSI) based on sequential hypothesis testing and Bayesian updating by Baheti et al. \cite{neifeld2008adaptive}.
        AFSI demonstrated that certain recognition tasks, such as face identification, could be performed without explicit image reconstruction.
        %
        Our work adopts this "reconstruction-free" philosophy but diverges in its optimization strategy.
        AFSI operates as an online, iterative system that requires real-time feedback, whereas our approach focuses on the offline optimization of static coding matrices, offering higher efficiency for high-speed hardware implementation.
        Despite these advances, a fundamental gap remains in the treatment of noise physics within the FSI framework. 
        FSI, including the AFSI variant,  has primarily been optimized under AGN assumptions \cite{neifeld2008adaptive, neifeld2014optimizing}.
        However, as acknowledged in \cite{neifeld2014optimizing}, a principled treatment of signal-dependent PN lies beyond the scope of classical linear formulations.
        Our work addresses this limitation by optimizing static, task-specific measurement masks directly under PN-limited conditions, thereby extending the FSI paradigm to photon-counting regimes.
    }
    \item \textbf{Noise-dependent Coding Optimization}. 
    Conventional works in CI area, \YLnote{largely reconstruction driven}, have assumed AGN as the primary noise source and recognized Hadamard as the best coding design \cite{cossairt2012does, harwit1979hadamard, wuttig2005optimal}. 
    It is recognized that coding optimal for AGN is suboptimal for PN in many different CI applications \cite{willet2009CSPoisson, harwit1979hadamard, cossairt2012does, mitra2014can, wuttig2005optimal}. 
    Harwit et al. \cite{harwit1979hadamard} proved the optimality of Hadamard coding under signal-independent noise, but opposed using it under PN. 
    Raginsky et al. \cite{willet2009CSPoisson} have proved the upper bound of the reconstruction error in compressed sensing behaves differently when transiting from AGN to PN.
    Cossairt et al. \cite{cossairt2012does} and Mitra et al. \cite{mitra2014can} have demonstrated that AGN-optimal coding designs cannot give rise to ideal performance gain when PN is predominant.
    Mitra et al. \cite{mitra2014can} further proposed that data-driven coding designs using data priors outperform classical ones such as Hadamard.
    Wuttig and Ratner et al. \cite{wuttig2005optimal, ratner2007optimal} introduced Poisson-noise-aware optimization techniques for coding matrices that outperform Hadamard in the presence of moderate signal-dependent noise. 
    However, their analysis neglects the data priors, leading to suboptimal matrices for greater Poisson noise \cite{mitra2014can}.
    Additionally, attempts at matrix optimization focusing on minimal mutual coherence \cite{mordechay2014matrixRIP} do not adequately consider Poisson noise, as highlighted in this study. 
    %
    %
    %
    \item \textbf{End-to-end Optimization and Learned Optics}. 
    This method jointly optimizes optical coding design and digital signal processing to achieve optimal overall performance \cite{diamond2021dirty, zhang2021deep, jacome2023middle, klinghoffer2022physics}. 
    In previous works, this idea was usually implemented without considering Poisson noise in cost functions \cite{chang2018hybrid, hinojosa2021learning, dun2020learned, metzler2020deep, chang2019deep, onzon2021neural, spall22hybrid_training,  gibson2020single, jacome2023middle} or without optimizing masks with gradient incorporating Poisson distribution \cite{diamond2021dirty, rego2022deep, duarte2008CS, nature2022ONN, gibson2020single, xu2020compressed}. 
    Chang et al. \cite{chang2018hybrid} proposed a hybrid optical-electronic model demonstrating learning optical elements. 
    Their primary goal is computational efficiency, and their system is not designed for Poisson noise. 
    Wang et al. \cite{nature2022ONN} successfully implemented a neural network model for handwritten number classification on an optical device with limited photon budget, demonstrating the potential for AI-assisted optimization of coding schemes.
    However, Poisson noise was considered only in model testing where the most robust model was picked among a set of {different initialization seeds} \cite{nature2022ONN}. 
    %
    Xu et al. \cite{xu2020compressed} examined the robustness of the proposed neural network model to Poisson noise, but their training process does not explicitly incorporate the gradients affected by Poisson noise.
    Tseng et al. \cite{tseng2021differentiable} proposed a noise-aware optimization methodology, but their work focuses on the point-spread function on camera design.
    Mitra et al. \cite{mitra2014can} used a Gaussian Mixture Model and priors from training data to optimize coding, but this work focuses on image reconstruction.
    \YLnote{
        %
        %
        Within reconstruction‑free learned‑optics for single‑pixel classification, Bacca et al. \cite{bacca2020coupled} optimize coded apertures for compressive classification. 
        We recognize that their choice of classification tasks provides a valuable benchmark for the community, and adopted similar tasks to evaluate our performance and their methodology as a reference point in our AGN-baseline.
        %
    }
    While prior works have explored end-to-end learning for various imaging tasks and some have addressed noise during inference, none have specifically extended FSI paradigm to jointly optimize the optical sensing matrix and feature extraction network explicitly for the challenging Poisson noise regime.
    Our work on DeepFSI directly addresses this critical gap arising from the insufficient understanding of Poisson noise and enhances the practicality of FSI techniques.

\end{itemize}


    


\section{Background}
\label{sec:3_background}

\subsection{Single-pixel Camera}

\begin{figure}[ht]
    \centering
    \begin{subfigure}{0.45\linewidth}
        \centering
        \caption{}
        \includegraphics[width=1\linewidth]{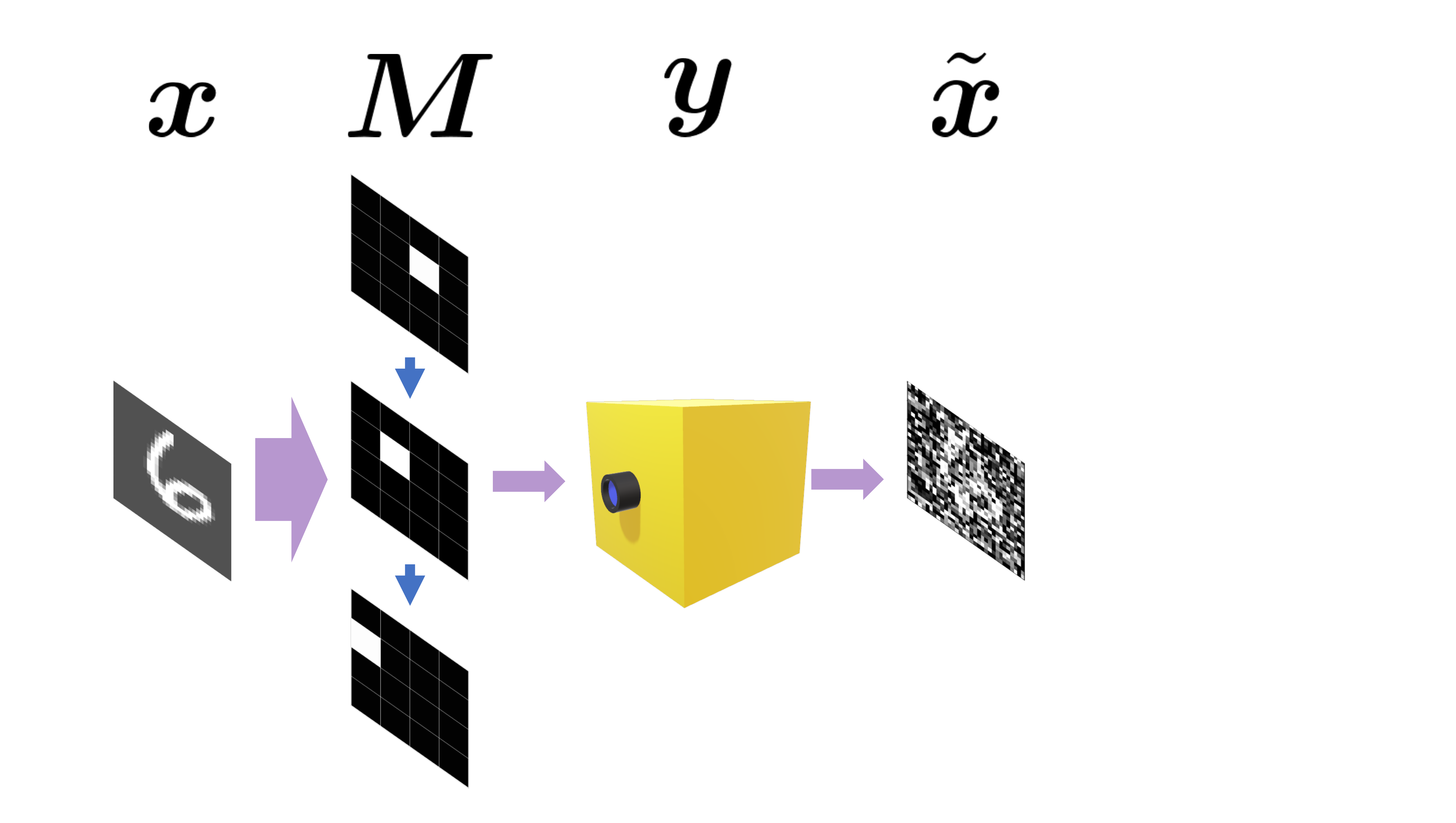}
        \label{fig:RasterScan}       
    \end{subfigure}
    \begin{subfigure}{0.45\linewidth}
        \centering    
        \caption{}
        \includegraphics[width=1\linewidth]{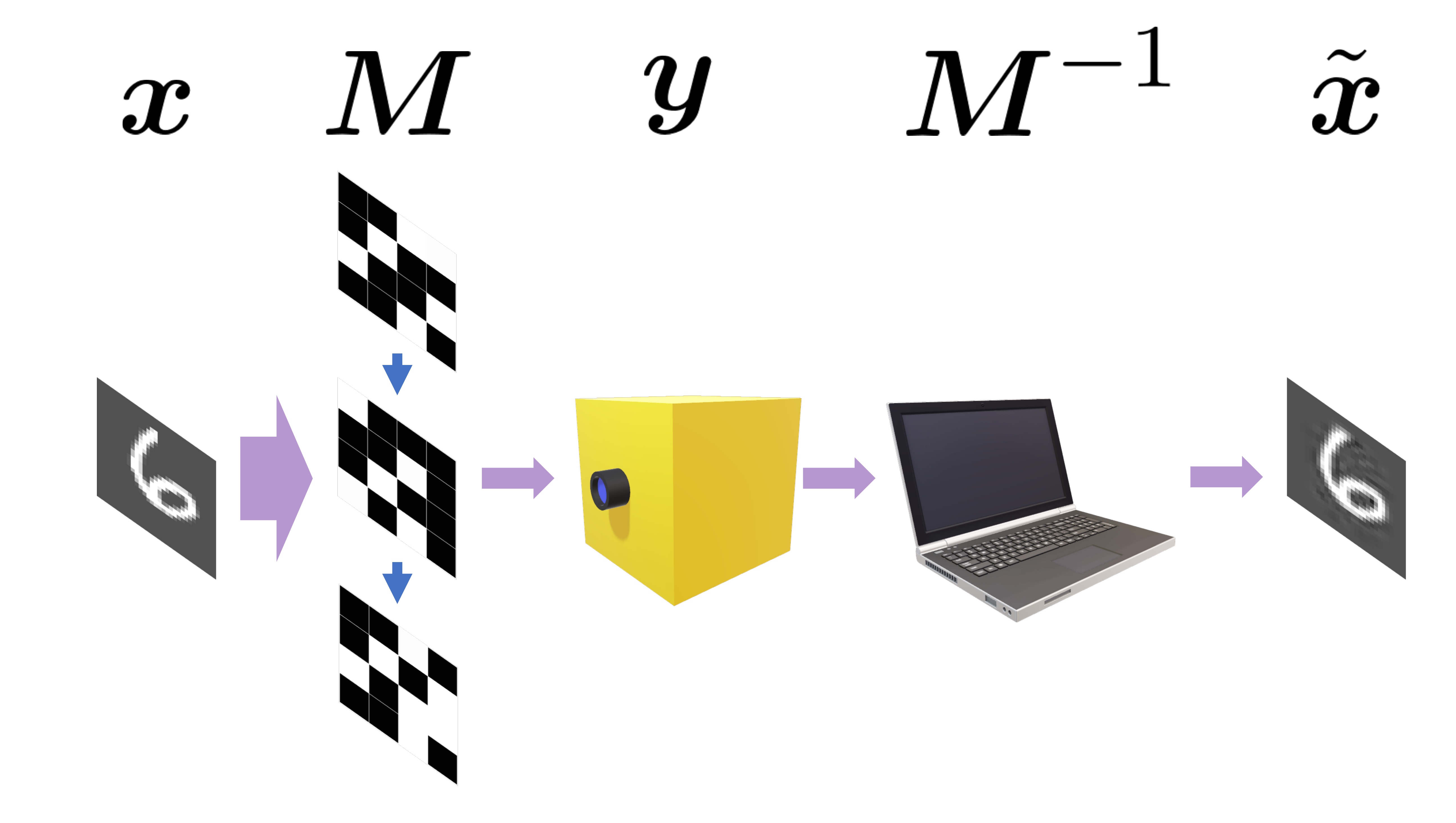}
        \label{fig:BasisScan}        
    \end{subfigure}
    \caption{
    Single-pixel camera configuration. 
    $\B{x}$: field of view, $\B{M}$: masks of coding, $\B{y}$: counted photon numbers at the sensor, $\B{M}^{-1}$: reconstruction operator, $\tilde{\B{x}}$: reconstructed object. (\subref{fig:RasterScan}) Only one pixel (white) is scanned in each measurement, and the measured data requires no reconstruction. (\subref{fig:BasisScan}) The sum of all white pixels is measured in each measurement and it requires a decoding step to reconstruct the field of view.
    }
    \label{fig:SinglePixelImaging}
\end{figure}

The single-pixel camera (SPC) shown in Fig. \ref{fig:SinglePixelImaging} is a popular example that demonstrates and tests the effect of a given coding design. 
Coding optimization focuses on the spatial dimensions, but the methodology obtained from it can be generalized to other coding-based vision systems. 
Our SPC uses a photon multiplier tube (PMT), a rapid single-pixel detector using photon-counting technology, in conjunction with a digital micromirror device (DMD) to sequentially project a sensing matrix $\B{M}$ by modulating various patterns on the DMD \cite{mitra2014can}. 
Its measurement process mathematically follows the Eq. \ref{eqn:NoiselessMeasurement}. 
Here, $\B{x} \in \IR^{N \times 1}$ is the image representation of the field of view (FOV), which consists of $N$ pixels, and is measured by linear projections or masks, $\B{M} \in \IR^{m \times N}$ to attain corresponding flux levels or photon counts, $\B{y} \in \IR^{m \times 1}$ 
 \cite{willet2009CSPoisson}. 
%
%
Physically, $\B{M}$ is the optical coding, and can be implemented on the DMD by directing or blocking different parts of the incoming light from $\B{x}$ \cite{raskar2009computational} and averaging them on sensors that digitize the detected flux levels, or photon counts, $\B{y}$. 
The patterns used for this objective are called masks, corresponding to the rows of $\B{M}$.
If the sensing matrix $\B{M}$ is full-rank, $\B{x}$ can be reconstructed by $\B{M}^{-1}\B{y}$. 
However, even when $m < N$, we can still recover $\B{x}$ via compressed sensing. 
Although Hadamard matrices are considered effective in coding, their suitability diminishes in the presence of data-dependent Poisson noise \cite{mitra2014can, harwit1979hadamard}.
%
%
In the FSI narrative, coding can also be interpreted as feature extraction, where $\B{y}$ can be used directly in classification tasks as a feature vector. 
In this case, the following recovery of $\B{x}$ in Fig. \ref{fig:SinglePixelImaging} is not necessary.

\subsection{Selection of Sensing Matrix}
In this work, we categorize different types of $\B{M}$ into raster scan, impulse imaging, non-FSI codes, and FSI codes. 
Raster scan (RS), demonstrated in Fig. \ref{fig:RasterScan}, refers to a pixel-wise scanning approach, while impulse imaging (II), the gold standard in this work, involves an ideal sensor collecting the entire data spectrum simultaneously.
In this work where the object $\B{x}$ is a 2D image, II can be interpreted as an N-pixel array and other sensing matrices remain on a single pixel.
However, when $\B{x}$ is \Xpolish{a hyperspectral}{hyperspectral data}, the fabrication of impulse imaging camera can be very challenging.
Since each pixel in II is exposed for the entire measurement period, it is explicit that 
\YLnote{II provides the upper bound among all
strategies in this work.}
Non-FSI and FSI codes constitute basis scan, shown in Fig. \ref{fig:BasisScan}.
Non-FSI codes include classic, task-independent masks such as Hadamard basis (HB)
as well as \YLnote{\Xpolish{truncated}{Truncated}} Hadamard (TH) codes, which preserve only the low-frequency components of the Hadamard matrix. 
FSI codes are task-specific and must be derived from training data, with PCA-based FSI and the proposed DeepFSI falling into this category.

\subsection{Model under Noise}
\label{ssec:model_inder_noise}

This project investigates two noise models. 
The first one is AGN related to dark current or read noise from imperfect sensor materials, which mainly originates from thermal vibrations of atoms at sensors. 
In this noise model, Eq. \ref{eqn:NoiselessMeasurement} becomes:
\Xequa{
\label{eqn:gaussian_measurement}
\noisy{\B{y}} = \B{Mx} + \B{\epsilon},
}
where $\B{\epsilon} \sim \AGN(\B{0}, \sigma^2\B{I})$, and $\sigma$ is the standard deviation. 
The other noise model is photon noise, which arises from the statistical nature of photons \cite{boyat2015review}. 
In most prior research, photon \YLnote{noise} is approximated as \YLnote{Poisson noise (PN)} 
and the measurement equations associated with this noise model typically follow a Poisson distribution:
\begin{equation}
\label{eqn:poisson_measurement}
\begin{aligned}
\noisy{\B{y}} &\sim \Poisson(\B{Mx})\\
\Pr(\noisy{{y}}_i &=  k) = \frac{y_i^k \exp(-y_i)}{k!}
\end{aligned} \quad,
\end{equation}
where $y_i = \sum_{j=1}^N M_{ij} x_j$ \cite{willet2009CSPoisson}.  
While PN is the best way to approximate photon noise at this stage, 
its nonlinearity adds complexities to the design and optimization of affected systems, and the use of Poisson random function eliminates gradients, making it difficult to optimize the hardware components in an end-to-end model.
To address these shortcomings, this project focuses on the signal dependency of PN and adopts
the MLGauss model that uses a Gaussian variable with \Xpolish{intensity-dependent variance}{mean-variance equivalence} \cite{selwood2022coded_aperture_imaging, mitra2014can, cossairt2012does}, formulated as

\begin{equation} \left\{
    \begin{aligned}
        \noisy{\B{y}} &= \B{Mx} + \B{J},\\
        \B{J} &=  (\B{Mx})^{\circ \frac{1}{2}} \odot \B{\epsilon},\\
        \B{\epsilon} &\sim \mathcal{N}(0, \B{I}),
    \end{aligned} \right. \quad 
\end{equation}
 where $\circ \frac{1}{2}$ represents element-wise square root, and $\odot$ stands for element-wise product. It substitutes conventional Poisson approximations whenever a gradient computation is required in training stage, yet the term "Poisson" is retained for consistency with other classical models involving no hardware optimization, facilitating straightforward comparisons. 
 %
 %
 For model testing, all models employ the PN model, as gradient computation is no longer required.

\subsection{Restrictions of Coding Designs}


While we can identify the optimal sensing matrix $\B{M}$ tailored to the tasks at hand, we acknowledge that the optimization problem about $\B{M}$ may not exhibit convexity. 
This is attributed to the presence of two constraining factors. 
Consequently, achieving globally optimal masks may not be universally feasible within the scope of this problem.

\begin{enumerate}
    \item\label{constraint:photonNumber} \textbf{Flux-preserving} \cite{willet2009CSPoisson}. The SPC model involves the allocation of photons available  within $\B{M}$, as discussed in \cite{neifeld2003dual}. 
    It is important to ensure that $\B{M}$ does not produce additional photons through improper entries \cite{neifeld2003dual}. 
    Mathematically, 
     $\sup \sum_{i=1}^m M_{ij} = 1, \forall j \in \{1,2,\dots,N\}$ \cite{neifeld2003dual}.
    %
    \item\label{constraint:negativeEntry} \textbf{Positivity-preserving} \cite{willet2009CSPoisson}. 
    It is not possible to physically implement negative values for masks $\B{M}$, as demonstrated in \cite{neifeld2003dual, willet2009CSPoisson}. 
    In this project, we used the dual-rail approach as outlined in the work of Neifeld et al. where positive and negative entries of a mask give rise to two separate measurements \cite{neifeld2003dual}.
    Mathematically, for the $i_\text{th}$ mask, $y_i = F(\sum_{j=1}^N \sgn{M_{ij}} M_{ij} x_j) -  F(-\sum_{j=1}^N\sgn{-M_{ij}} M_{ij} x_j)$, where operator $F$ is about the noise generation explained in Section \ref{ssec:model_inder_noise} and $\text{sgn}$ is the step function returning $1$ for positive values and $0$ for negative values. 
    %
    
\end{enumerate}

\section{Methods}
\label{sec:4_methods}
\subsection{Model Design}

\begin{figure}[ht]
    \centering
    \begin{subfigure}{\linewidth}\centering
    \caption{\YLnote{Scanner-Classifier Configuration}}
    \includegraphics[width=0.9\linewidth]{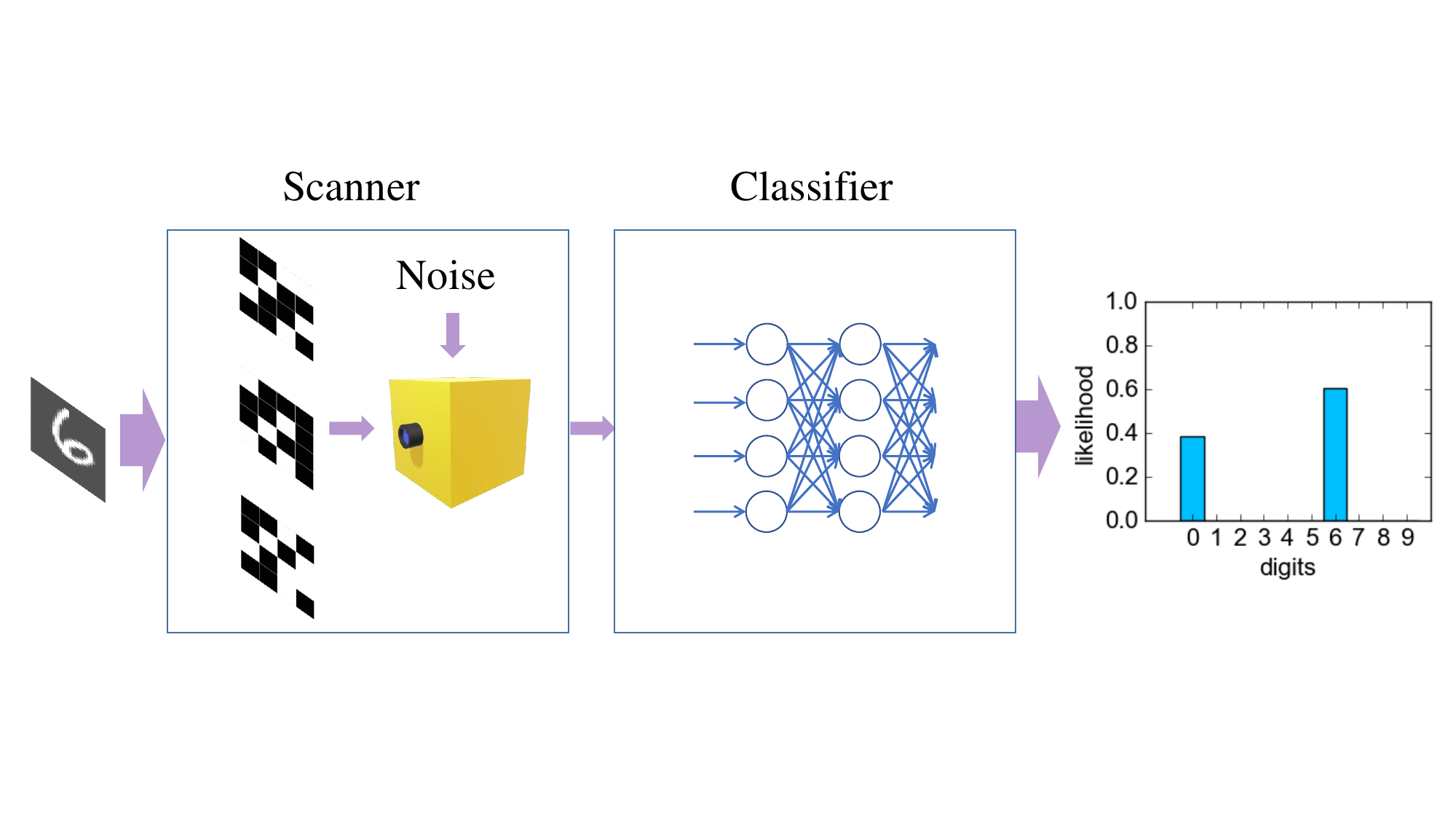} 
    \label{fig:model_config_a}
    \end{subfigure}

    \begin{subfigure}{0.7\linewidth}\centering    
    \caption{Graph Representation}    

\newcommand{\Wd}{0.12\linewidth}
\newcommand{\Ceil}{1.8}
\newcommand{\Floor}{-1.8}
\newcommand{\Ht}{0}


\tikzset{%
  every input_neuron/.style={
    circle,
    draw = black
    },    
  every optical_neuron/.style={
    diamond,
    fill=green,
    draw = black,
    minimum size=0.8cm
    },
  every neuron/.style={
    circle,
    draw,
    minimum size=0.5cm,
    fill=cyan
  },
  neuron missing/.style={
    draw=none, 
    fill=none,
    scale=1,
    text height=0.5cm,
    execute at begin node=\color{black}$\vdots$
  },
  every noise/.style={
    shape=diamond,
    fill=green,
    draw=black,
    minimum size=0.8cm
  }
}



\begin{tikzpicture}[x=1cm, y=1cm, >=stealth]

\foreach \m/\l [count=\y from 0] in {1,2,missing,3,4}
  \node [every input_neuron/.try, neuron \m/.try] (input-\m) at (0*\Wd,\Ceil - \y * \Ceil / 4 + \y * \Floor / 4) {};
  
\foreach \m [count=\y from 0] in {1,2,missing,3,4}
  \node [every optical_neuron/.try, neuron \m/.try] (DMD-\m) at (1*\Wd,\Ceil - \y * \Ceil / 4 + \y * \Floor / 4) {};

\foreach \m [count=\y from 0] in {1,2,missing,3,4}
  \node [every noise/.try, neuron \m/.try] (noise-\m) at (2*\Wd,\Ceil - \y * \Ceil / 4 + \y * \Floor / 4) {};
  
\foreach \m [count=\y from 0] in {1,2,missing,3,4}
  \node [every neuron/.try, neuron \m/.try ] (hidden1-\m) at (3*\Wd,\Ceil - \y * \Ceil / 4 + \y * \Floor / 4) {};

\foreach \m [count=\y from 0] in {1,2,missing,3,4}
  \node [every neuron/.try, neuron \m/.try ] (hidden2-\m) at (4*\Wd,\Ceil - \y * \Ceil / 4 + \y * \Floor / 4) {};


\foreach \m [count=\y from 0] in {1,2,missing,3,4}
  \node [every neuron/.try, neuron \m/.try ] (output-\m) at (5*\Wd,\Ceil - \y * \Ceil / 4 + \y * \Floor / 4) {};


\foreach \l [count=\i] in {1,2,n-1,n}
  \draw [<-] (input-\i) -- ++(-1,0)
    node [above, midway] {$X_{\l}$};

\foreach \l [count=\i] in {1,2,m-1,m}
  \node at (DMD-\i) {\textcolor{red}{\tiny${\l}$}};
\foreach \l [count=\i] in {1,2,m-1,m}
  \node at (noise-\i) {\textcolor{red}{\tiny${\l}$}};
\foreach \l [count=\i] in {1,2,39,40}
  \node at (hidden1-\i) {\tiny${\l}$};
\foreach \l [count=\i] in {1,2,127,128}
  \node at (hidden2-\i) {\tiny${\l}$};

\foreach \l [count=\i] in {1,2,9,10} 
  \draw [->] (output-\i) -- ++(1,0)
    node [above, midway] {$O_{\l}$};

\foreach \i in {1,...,4}
  \foreach \j in {1,...,4}
    \draw [->] (input-\i) -- (DMD-\j);
    
\foreach \i in {1,...,4}
    \draw [->] (DMD-\i) -- (noise-\i);

\foreach \i in {1,...,4}
  \foreach \j in {1,...,4}
    \draw [->] (noise-\i) -- (hidden1-\j);

\foreach \i in {1,...,4}
  \foreach \j in {1,...,4}
    \draw [->] (hidden1-\i) -- (hidden2-\j);

\foreach \i in {1,...,4}
  \foreach \j in {1,...,4}
    \draw [->] (hidden2-\i) -- (output-\j);

\foreach \l [count=\x from 0] in {Input, $\B{M}$, Noise,  ,  , Output}
  \node [align=center, above] at (\x*\Wd,\Ceil+0.5) {\small\l};

  \node [align=center, above] at (3.5*\Wd,\Ceil+0.5) {\small{Fully connected} \\ \small{layers}};

\end{tikzpicture}



    \label{fig:model_config_b}
    \end{subfigure}
    \caption{(\subref{fig:model_config_a}) The general configuration of scanner-classifier networks. 
    (\subref{fig:model_config_b}) The node-level architecture of our scanner (green diamonds)-classifier (blue circles) network. 
    Noise is implemented after the sensing matrix $\B{M}$. 
    Different from other coding schemes, DeepFSI has a trainable scanner where $\B{M}$ is not fixed and can be optimized by the gradient of the classifier.}
    \label{fig:model_config}
\end{figure}

Classification tasks are the proper venue to test feature fidelity of coding designs. 
All coding designs in this project for classification tasks share the same fundamental model structure, as illustrated in Fig. \ref{fig:model_config}. 
In this project, we apply our model to the classification of handwritten number with the MNIST dataset.
%
To classify handwritten numbers using optical devices with limited photon budget, we built a PyTorch-based pipeline consisting of a hardware-end \textit{scanner} module including $\B{M}$ and a noise generator to simulate the acquisition of $\noisy{\B{y}}$ described in Eq. \ref{eqn:gaussian_measurement} and Eq. \ref{eqn:poisson_measurement}, and a software-end \textit{classifier} module classifying the measured objects by $\noisy{\B{y}}$.
For the classification task, we utilized a two-hidden layer artificial neural network (ANN) with 40 and 128 nodes, serving as the primary model. 
To avoid overfitting and enhance generalizability, we applied dropout after each hidden layer, followed by a rectified linear unit (ReLU) activation layer.
During model training, the scanner computes the photon counts with noise and then transmits these data to the classifier. 
The noise is sampled from a predefined noise distribution and varies across epochs. 
Once the loss has been computed, the gradient descent is used to optimize the classifier parameters. 
%
What sets DeepFSI apart from other coding designs is the learnable $\B{M}$ that is not predefined nor fixed. 
In other words, backpropagation is extended to the scanner, allowing the simultaneous optimization of the scanner and the classifier under different noise models. 
We adopt the reparameterization trick, which is incorporated in PyTorch, to estimate the gradient affected by noise simulated by random number generator functions.
This approach offers distinct advantages compared to traditional methods, which typically optimize the classifier independently of the scanner or optimize the scanner without the appropriate noise model.
In our design, though the hybrid optimization fully 
\YLnote{\Xpolish{replies}{relies}} on computer simulation, it is feasible to load the scanner parameters into the coding optics to build a real camera, which is also demonstrated in this work.
Once operational, this model can serve various purposes, ranging from object classification and signal recovery to end-to-end camera design.
%
%
%

\begin{figure}[ht]
    \centering    
        \includegraphics[width=.75\linewidth]{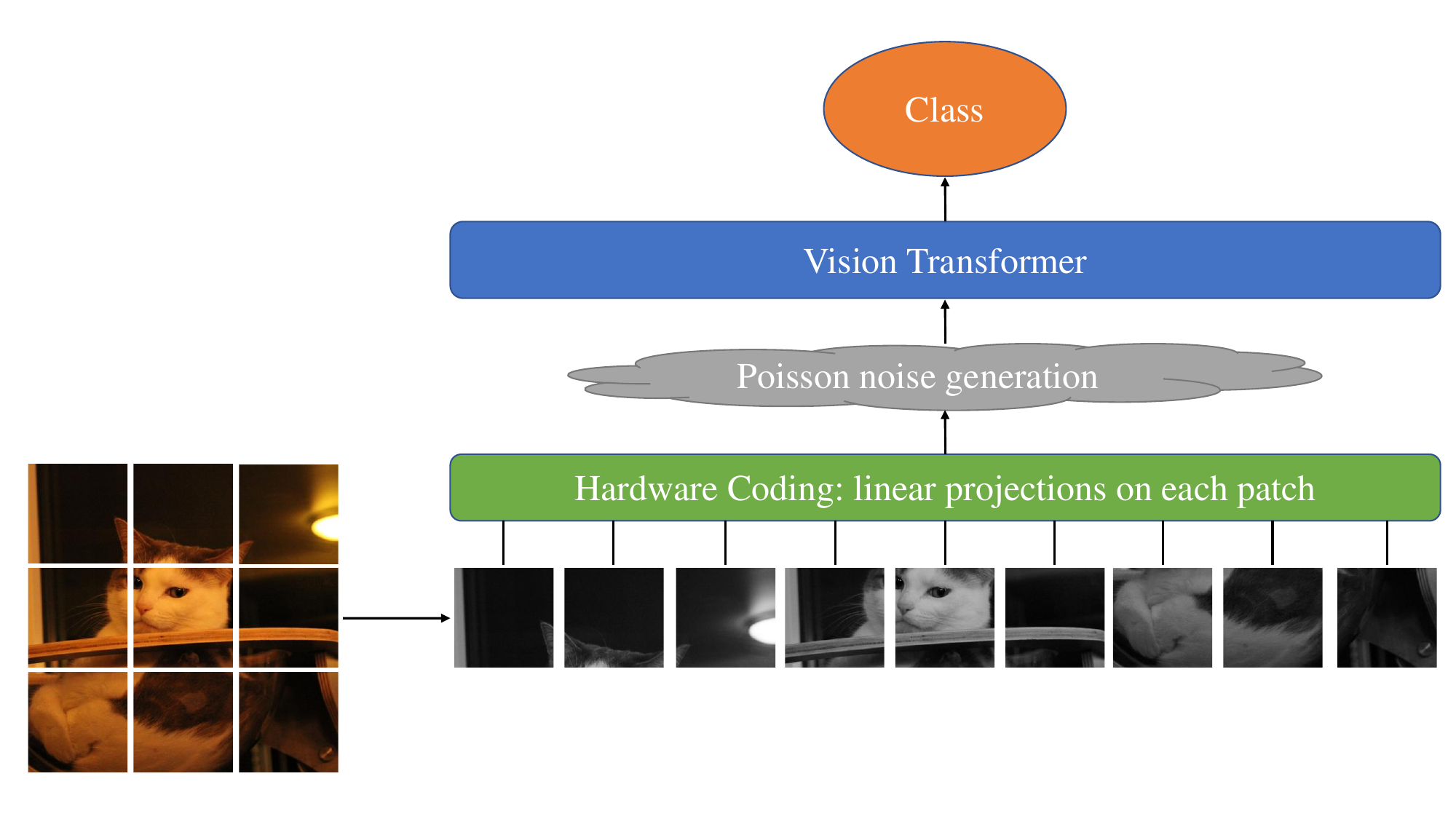}   
    
        \caption{
        OViT Overview. 
        The process begins by dividing the input image into fixed-size grayscale patches. 
        Poisson noise is then added after the initial linear projection to mimic optical coding within a vision system. 
        These noisy embeddings are subsequently fed into a Vision Transformer.  
        The picture is photographed by the author (Lu).
        }
    \label{fig:OVit_overview}    
\end{figure}

To further prove the necessity of our hardware optimization methodology and its versatility across a wide range of modern computer vision tasks beyond MNIST, we use the same scanner model with a Vision Transformer (ViT) as the classifier. 
The ViT model is adjusted to work with CIFAR-10 data resized to 64 by 64 pixels.
To better simulate a real computer vision system, we use only grayscale images with smaller 8-by-8 patch sizes following the classical FSI configuration \cite{neifeld2003FSI}, without any image enhancements or other preprocessing steps.
The overall architecture is shown in Fig. \ref{fig:OVit_overview}. 
This model is referred to as the Optical-frontend Vision Transformer (OViT) in this context, and it serves as another DeepFSI example.

\subsection{Simulated Experiments Design}
\label{ssec:simulated_experiments}

\YLnote{
    In this section, we present three classification tasks to explore the upper bound performance of the proposed framework under various hyperparameter configurations, following a standard vision system implementation workflow that includes simulation-based optimization and real-world deployment. 
    The hardware tests, presented in the following section, evaluate the transfer robustness of the optimized design as an integral part of this workflow. 
}
Initially, we evaluated the effectiveness of various coding schemes using the MNIST dataset, focusing on their performance in number recognition. 
Subsequently, we extend the evaluation to
CIFAR-10 by OViT
, examining the adaptability and efficacy of DeepFSI in this different domain. 
\YLnote{We also test the proposed method on hyperspectral data to see its performance on more realistic scenarios.}
The MNIST dataset, which consists of handwritten digits ranging from 0 to 9, has a default size of 28 by 28 pixels \cite{lecun1998MNIST}. 
To align the data with Hadamard matrices, we add black pixels to the edges of the images and resize them to 32 by 32 pixels \cite{lecun1998MNIST}. 
%
Following this,
we rescale the pixel values from the original range of $[0,1]$ to $[0.3, 1]$. 
This rescaling introduces non-zero Poisson noise to the black regions, which has variance proportional to the expected photon counts.

We evaluate our proposed model through simulations under both AGN and Poisson noise models, using varying 
number of photons sent to the system 
to generate results at different noise levels. 
We also test all compressible strategies (TH, FSI, DeepFSI), varying the total number of masks $m$ 
for a given 
number of photons, and define a metric called the compression ratio as the ratio of the number of masks to the number of pixels, i.e. $m/N$ \cite{stojek2022define_compression_ratio}. 
For II, RS and HB, this ratio is restricted to 1.00, while the rest, TH, FSI, DeepFSI,  are evaluated over values in 
\{1\%, 4\%, 9\%, 16\%, 25\%, 36\%, 49\%, 64\%, 81\%, 100\%\}
so that $\B{M}$ of each coding scheme is not restricted to being full-rank. 
%
%
The final result for a certain number of photons and a certain compression ratio is the average of five independent trials.
As we \Xpolish{target}{focus} on investigating the maximum feature fidelity of each coding design, the best performance among the compression list is selected for final comparison.
%
%
We notice the initial $\B{M}$ for DeepFSI can affect performance in the presence of noise.
We therefore initialized its $\B{M}$ with the PCA components from the training data for a direct comparison with FSI.



On top of the classification tasks on MNIST dataset, we also try to apply DeepFSI to a Vision Transformer, i.e. OViT in this work, to show that this methodology can be applied on more challenging and complicated vision tasks, with modern computer vision models. 
The optimization of OViT is similar to the model used on MNIST classification.
However, no coding designs are compressed under the assumption that we have no prior knowledge about the optimal compression ratio, which is also  unavailable beforehand in most real-world projects. 
This work can serve as a good example that demonstrates the robustness of DeepFSI over traditional FSI, especially when the best compression ratio is unavailable or the test conditions are not similar enough.
%

\begin{figure}[ht]
    \centering
    \begin{subfigure}{0.52\linewidth}
        \centering      
        \caption{Indian Pines dataset \cite{PURR1947}} 
        \includegraphics[width=1\linewidth]{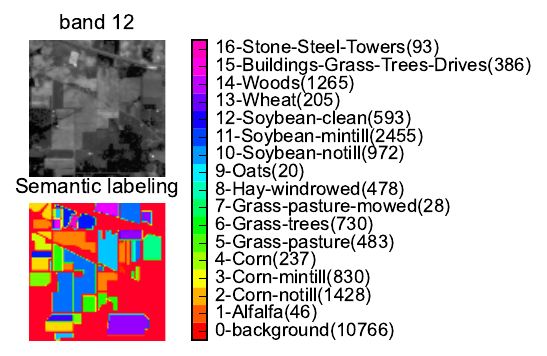}
        
        \label{fig:pines_dataset}    
    \end{subfigure}
    \begin{subfigure}{0.45\linewidth}
        \centering 
        \caption{Visualization of hyperspectrum examples}
        \includegraphics[width=1\linewidth]{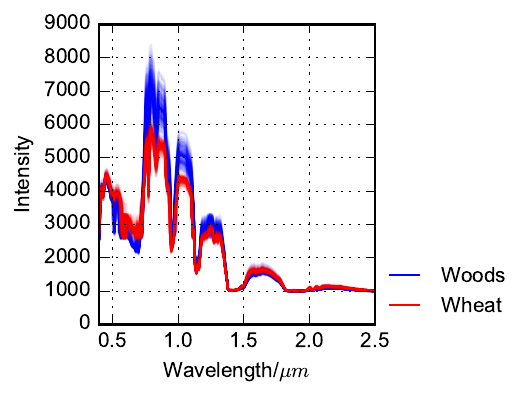}
        
        \label{fig:AeroRIT_Histogram}       
    \end{subfigure}
    \caption{\YLnote{IndianPines dataset overview}}
\end{figure}

\YLnote{
To further evaluate the generalization capability of DeepFSI beyond conventional image classification benchmarks, we conduct an additional experiment on hyperspectral data, which better reflects real-world sensing applications.
In this experiment, we \Xpolish{utilized}{utilize} the Indian Pines dataset \cite{PURR1947}, which was acquired through the Airborne Visible/Infrared Imaging Spectrometer (AVIRIS) on June 12, 1992, covering the Purdue University Agronomy farm northwest of West Lafayette and its surrounding area. 
%
This dataset consists of $145 \times 145$ pixels with 224 spectral reflectance bands spanning wavelengths from 0.4 to 2.5 $\mu m$, providing rich spectral information for material classification.
%
The goal is to assign each pixel's spectral signature to a material category.
%
\Xpolish{Fig.}{Figure} 
\ref{fig:pines_dataset} 
illustrates a representative spectral band (band 12), along with the class distribution within the dataset and the corresponding ground-truth semantic labels.
%
%
\Xpolish{Fig.}{Figure}  
\ref{fig:AeroRIT_Histogram} further shows the pixel-wise spectral responses for two representative classes across the 224 channels. 
%
Unlike standard image classification, feature extraction in this experiment is performed in the spectral (wavelength) domain rather than the spatial domain.
%
Spectral channels \Xpolish{were}{are} zero-padded from 224 to 256 to align with the Hadamard mask and rescaled to the range $[0,1]$. 
We \Xpolish{utilized}{utilize} the scanner-classifier module from Fig. \ref{fig:model_config}, adapting the \Xpolish{ONN}{DeepFSI} classification layers to the new dataset sizes. 
Each classification \Xpolish{ran}{run} for 2000 epochs, with a learning rate of $5 \times 10^{-3}$ and a batch size of 5000. 
%
Classification performances are evaluated cross varying \Xpolish{light levels}{photon budgets} from $10^1$ to $10^{10}$. 
}

\subsection{Hardware Experiments Design}



\begin{figure}[ht]
    \centering    
        \includegraphics[width=.6\linewidth]{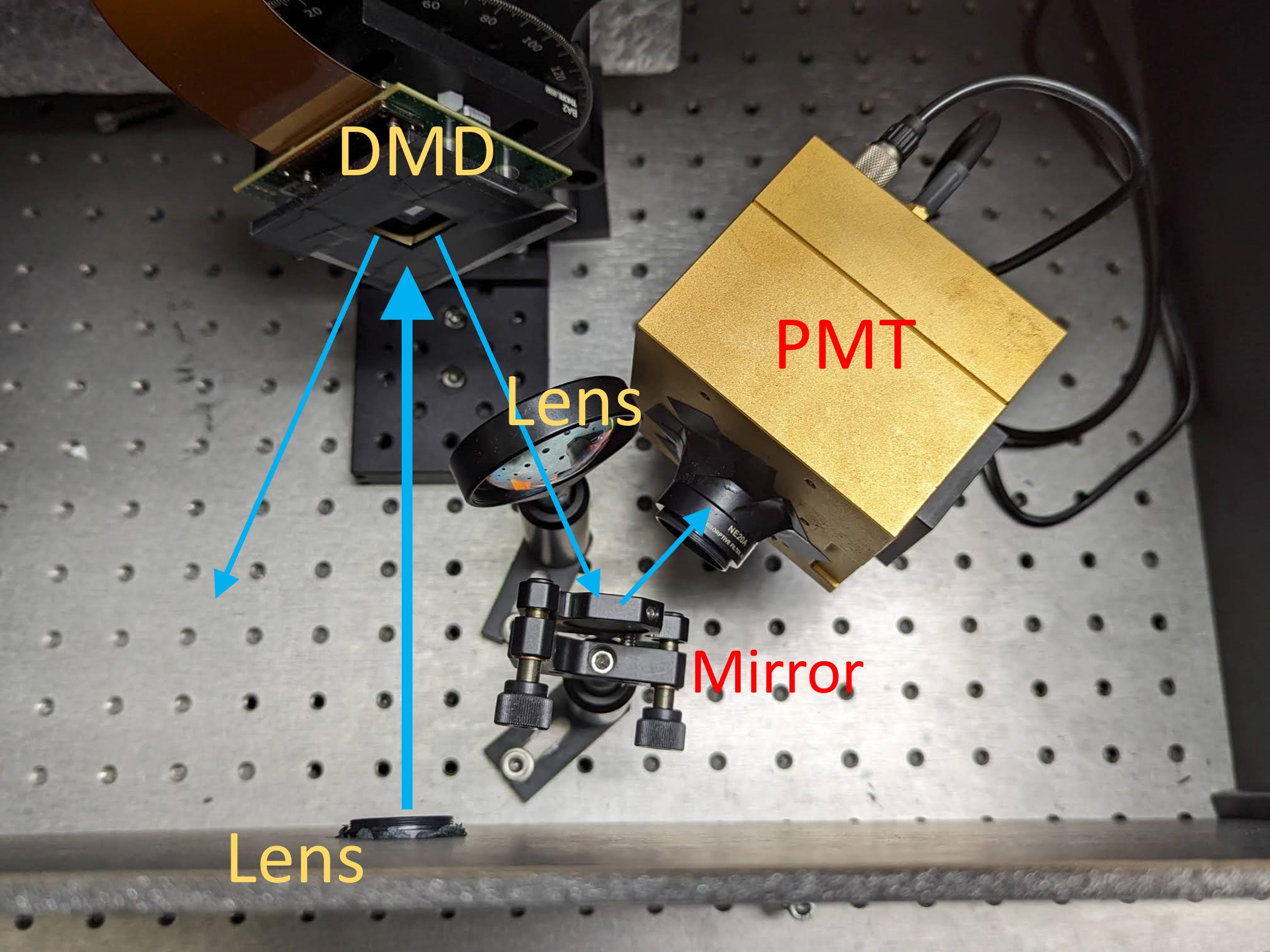}  
        %
    \caption{Experimental configuration of the single-pixel camera: The light path is illustrated by the blue arrows in the diagram. Specifically, our setup utilizes only one branch reflected by the DMD.}
    \label{fig:hardware_setup}
\end{figure}

A key concern with DeepFSI is that its optimization relies on simulation, and its \YLnote{feasibility in} real-world vision systems remains to be experimentally validated. 
%
\YLnote{
    In the simulated experiments of Section \ref{ssec:simulated_experiments}, the objective is to explore the theoretical performance upper bound of the proposed framework across a wide range of hyperparameters under controlled and idealized conditions.
    However, practical deployment inevitably introduces deviations from simulation assumptions. 
    Therefore, the primary goal of the hardware experiment is not to further improve peak accuracy, but to evaluate the transfer robustness of DeepFSI from simulation to a physical system.
    In particular, we investigate whether DeepFSI remains robust and generalized compared to other methods when experimental conditions, such as photon budget, optimal compression ratio, device bit depth, and ambient illumination, differ from those assumed during simulation and may be partially unpredictable.
}
%
To this end, we deploy the methodology developed for MNIST in Section \ref{ssec:simulated_experiments} on a laboratory SPC system.
%
As shown in Fig. \ref{fig:hardware_setup}, the setup consists of a DMD (digital micromirror device), specifically the DLP7000 model from Texas Instruments, and a PMT (photomultiplier tube), specifically the PMA Series from PicoQuant.
%
%
The training stage is conducted on a computer, and the parameters are deployed on the SPC to recognize handwritten \Xpolish{number}{digit} pictures displayed on a monitor.
The testset consisted of ten pictures and each handwritten number appeared exactly once
\YLnote{, but each digit is independently sampled at least 10 times to compute the average classification rates.}
%

\YLnote{
Similar to OViT experiments introduced in Section \ref{ssec:simulated_experiments}, the optimal compression ratio is not available.
We decide to use 0.09 in this case as its overall performance in the simulation is satisfactory.
In other word TH, FSI and DeepFSI use 92 masks,  while others use 1024 masks.
%
%
%
For each coding design, we test its performance when the total exposure times are  \Xhide{0.001, 0.01, 0.1, and 1.0}{0.0001, 0.001, 0.01, and 0.1}  seconds.
}

\begin{figure}[ht]
    \centering
    \begin{tabular}{>{\centering\arraybackslash}m{0.3\linewidth}>{\centering\arraybackslash}m{0.1\linewidth}>{\centering\arraybackslash}m{0.3\linewidth}}
         \includegraphics[width=1\linewidth]{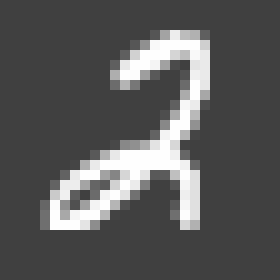}&
        $\longrightarrow$&
        \includegraphics[width=1\linewidth]{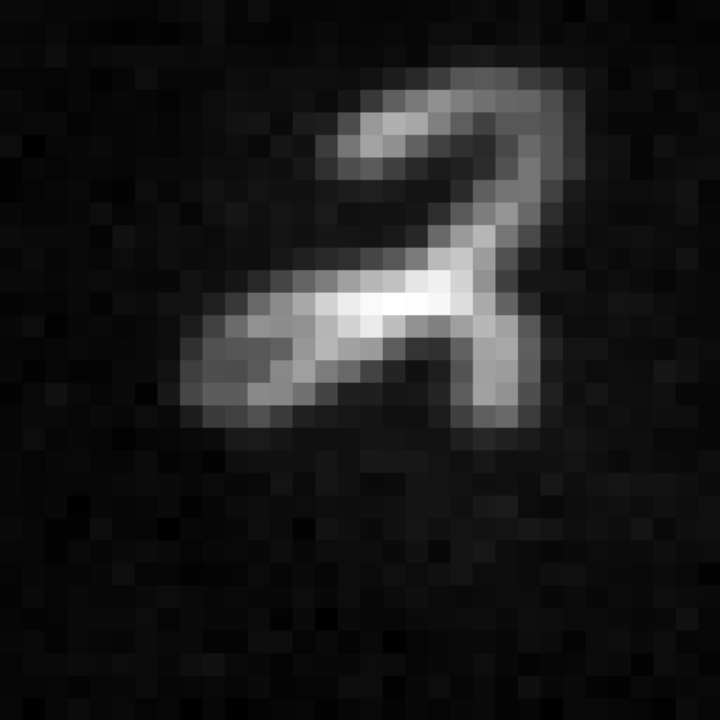}\\
    \end{tabular}
        
    \caption{Raw image 
    (left) VS SPC-observed image 
    }
    \label{fig:sensor_perception}
\end{figure}

\Xpolish{Fig.}{Figure}  
\ref{fig:hardware_setup} shows the hardware setup in this work.
%
The first crucial step is to transform the digital MNIST data to ensure that they are consistent with the pictures on the monitors in the SPC's perspective, as demonstrated in Fig. \ref{fig:sensor_perception}.
%
The transformed MNIST is then used in DeepFSI training, and the optimized mask-set digitalized from $\B{M}$ is loaded onto the DMD.
After the photon numbers are acquired, they are digitalized and vectorized and inputted to the classifier module on computer for number recognition.

\begin{figure}[ht]
    \centering
    \includegraphics[width=.5\linewidth]{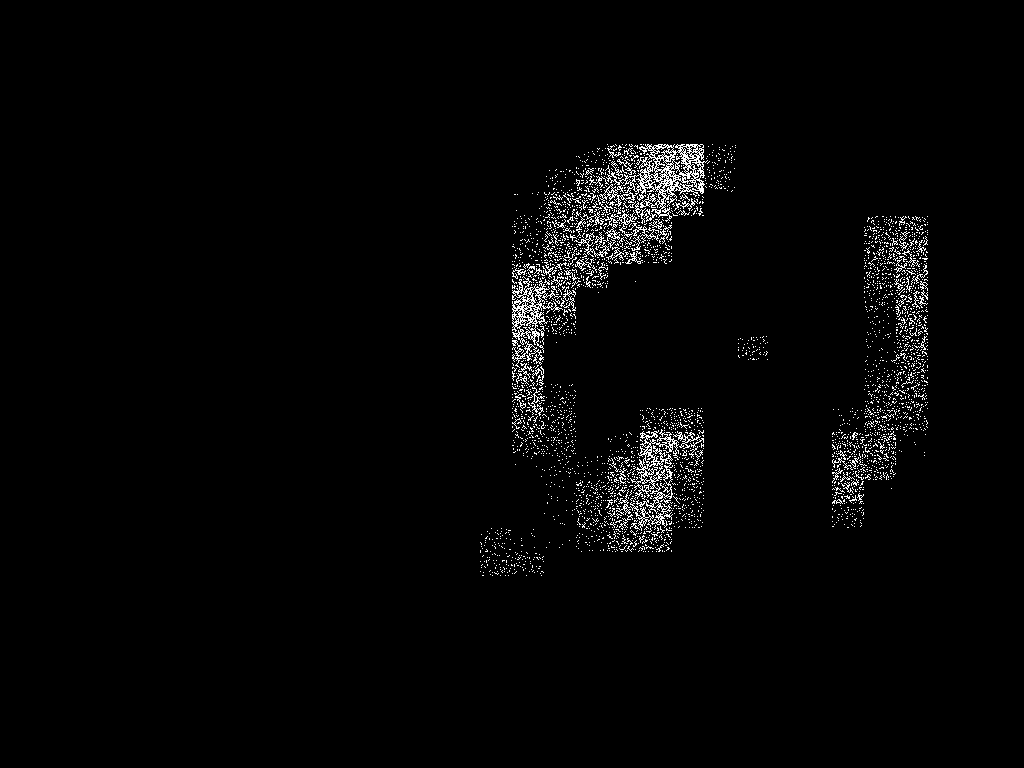}
    \caption{
    A DMD mask with decimal values. Black regions are OFF-state and white regions are ON-state mirror chips. A coding pixel consists of $32 \times 24$ mirror chips and the ratio of the white ones inside it represents the mask value at that pixel.
    }
    \label{fig:decimal_values_DMD}
\end{figure}

The mask-set is not required to be binary in this work, though DMD only have ON and OFF states. 
Our approach is to select a random fraction of micromirrors in a pixel to achieve decimal real numbers. 
The DLP 7000 consists of $1024 \times 768$ micromirrors, which project $32 \times 32$ masks. 
Therefore, one pixel includes $32 \times 24$ micromirrors.
Using this method, we can achieve a precision of $1.3\times 10^{-3}$, which is sufficient in this work.
%
\Xpolish{Fig.}{Figure}   
\ref{fig:decimal_values_DMD} shows an example of this trick.

\section{Results}
\label{sec:5_results}

\subsection{Simulated Experiments on MNIST}

\begin{figure}[hbt!]
    \centering
    \begin{subfigure}{.49\linewidth}
        \centering
        \caption{Classification under AGN}
        \includegraphics[width=1\linewidth]{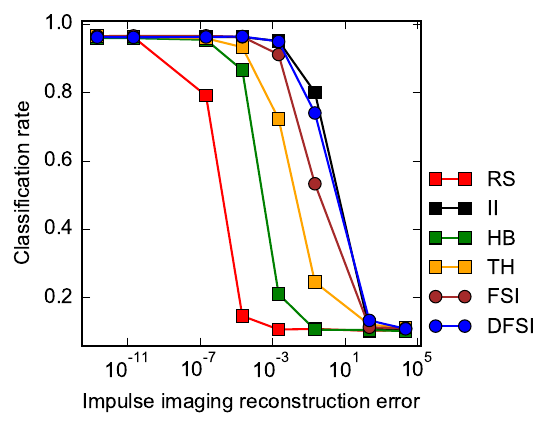}
        \label{fig:gaussian_classification}
    \end{subfigure}
    \begin{subfigure}{.49\linewidth}
        \centering
        \caption{Classification under PN}
        \includegraphics[width=1\linewidth]{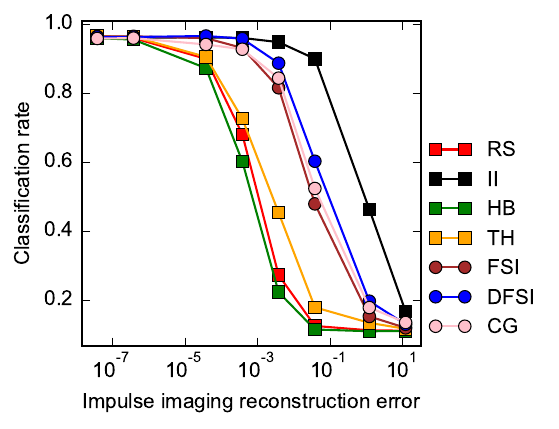}
        \label{fig:poisson_classification}
    \end{subfigure}
    \caption{Classification rates on \textbf{simulated data}. 
    RS: Raster scan. 
    HB: Hadamard basis. 
    II: Impulse imaging. 
    %
    %
    TH: Low Frequency Truncated Hadamard basis.
    FSI:  PCA basis. 
    DFSI: DeepFSI. 
    CG: control group. 
    Feature-specific methods are marked by $\bigcirc$ from traditional ones marked by $\square$.
    %
    The CG employs the optimized masks under the AGN and trains the software classifier under PN. 
    Its overall performance is still worse than DeepFSI.}
    \label{fig:simulated_classification}
\end{figure}

\Xpolish{Fig.}{Figure}   
\ref{fig:simulated_classification} displays the classification performance of different coding schemes in the simulated experiments. 
Since PN is signal dependent and AGN is constant in this work, comparing the performance under both models is not straightforward if we simply use the total number of photons as x-coordinates.
In this case, data points with the same x-coordinate are not affected by the same amount of noise when comparing across noise models.
In other words, when provided with a certain number of photons, the variances caused by AGN and PN are not equal. 
However, the reconstruction error of II is unique under an arbitrary photon number.
Therefore, we choose a metric of the reconstruction error with II to demonstrate the amount of noise regardless of its type and use it as the x-coordinate in the figure.
II is the gold standard, labeling the supreme among all the coding designs in this work.
By this metric, we can compare feature fidelity among different coding designs even between AGN and PN as long as the x-coordinates are the same.
%
As mentioned in Section \ref{ssec:simulated_experiments}, the simulations cover various compression ratios. To compare the maximum capacity of preserving features, we choose the best within different compression ratios for each coding design.
If a coding design outperforms another in classification in Fig. \ref{fig:simulated_classification}, we can identify that it has better feature fidelity \textbf{if used optimally}.
%
Feature-specific ones are indicated by circles, while non-feature-specific ones are marked by squares across both subfigures for clarity.

In Fig. \ref{fig:gaussian_classification}, we can identify that FSI and DeepFSI are much better than non-FSI coding designs. 
DeepFSI is also better than FSI and provides a performance most similar to that of the ideal II camera. 
All basis scan methods, including HB and TH that are without prior information, are better than RS, which scans pixels in sequence. 
This result aligns with the common sense that coding can improve the performance under AGN. 

In Fig. \ref{fig:poisson_classification}, we see that PN degrades the performance of any coding designs, and using non-FSI coding cannot provide a performance gain compared to RS. 
FSI presents better performance than non-FSI methods, although it is not designed for PN, which is consistent with the discovery of Mahalanobis et al. \cite{neifeld2014optimizing}.
It highlights the necessity of using data priors under PN in a vision system.
More significantly, DeepFSI outperforms traditional FSI, showing the advantage of using end-to-end optimization compared to PCA under PN.
Additionally, we present a control group that first optimizes the coding under AGN and subsequently optimizes the classifier with this coding fixed under PN.
We observe a suboptimal performance on the control group, indicating that using the optimal coding under AGN cannot lead to optimal performance in a vision system. 
It is an important discovery that the crucial role of noise models for the regime of future camera design. 
%


%
%
\begin{figure*}[htb!]
\footnotesize	
\centering
\begin{tabular}{>{\centering\arraybackslash}p{0.06\linewidth}|>{\centering\arraybackslash}m{0.27\linewidth}|>{\centering\arraybackslash}m{0.27\linewidth}|>
{\centering\arraybackslash}m{0.27\linewidth}
}
{Photons}  &  Gaussian & Poisson & Value Distributions\\
\hline
$10^3$ & \includegraphics[width=1 \linewidth]{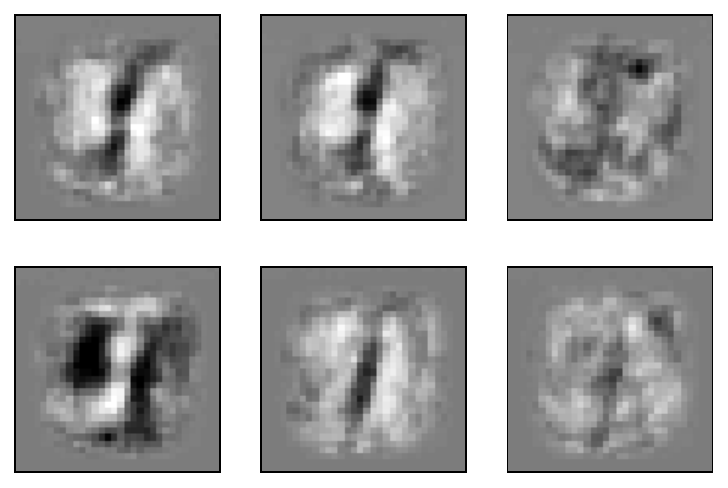} & 
\includegraphics[width=1\linewidth]{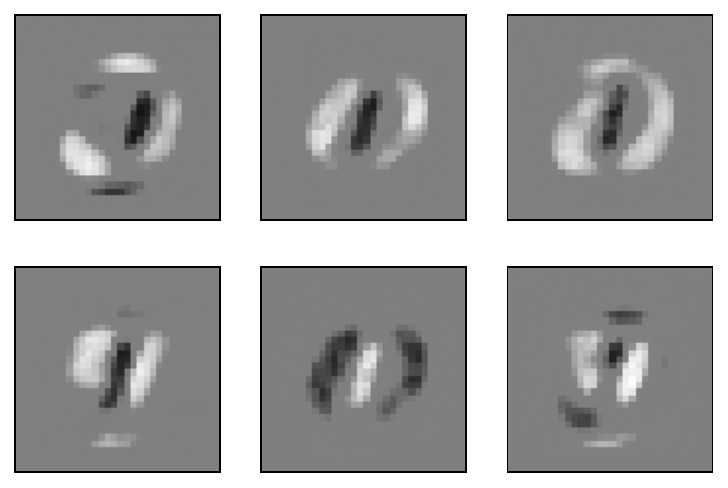} &
\includegraphics[width=\linewidth]{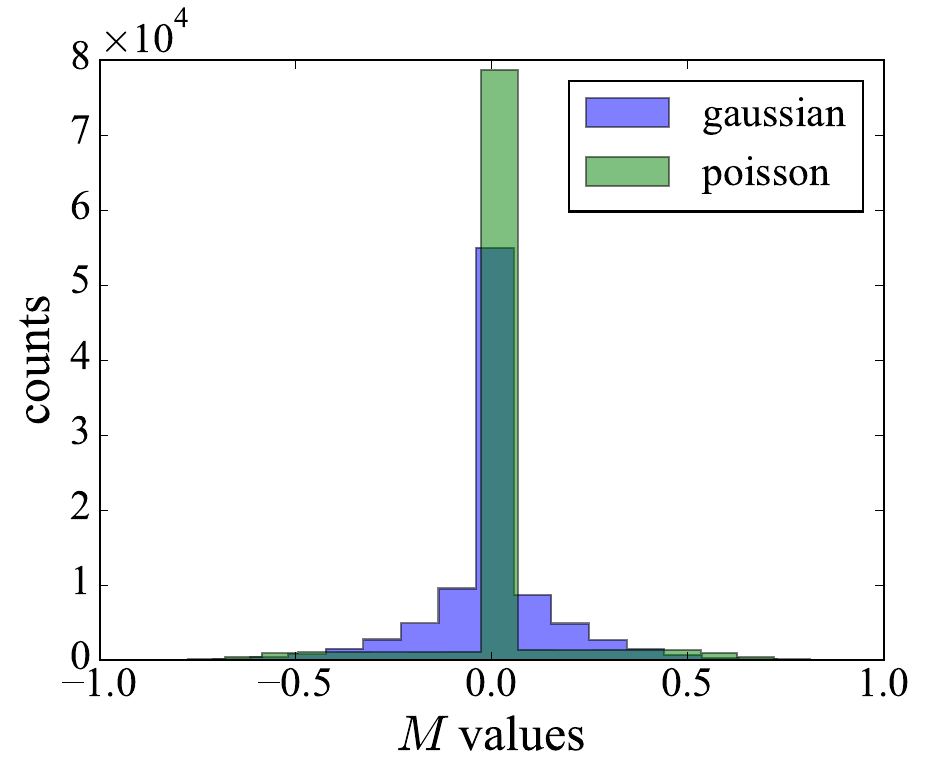}
\\
\hline
$10^5$ & \includegraphics[width=1 \linewidth]{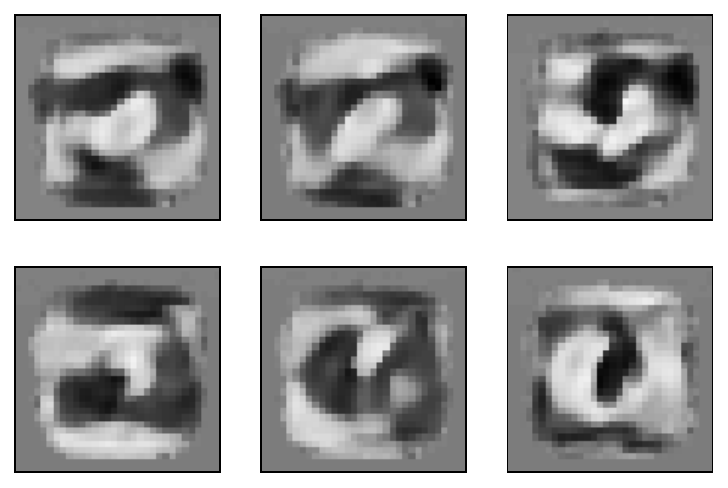} & 
\includegraphics[width=1\linewidth]{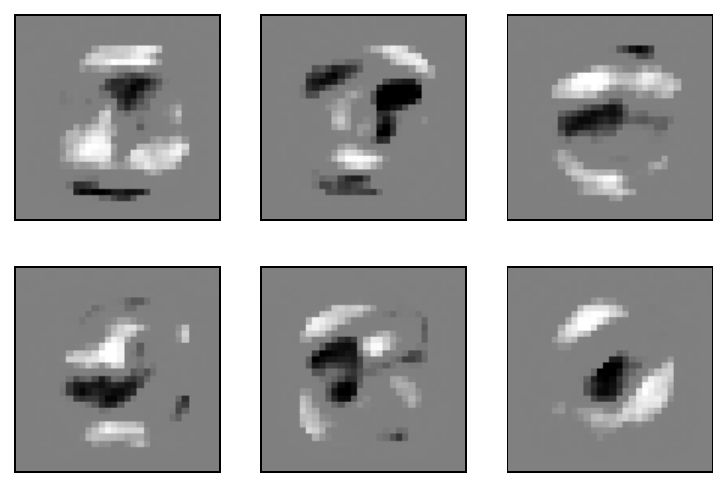} &
\includegraphics[width=1\linewidth]{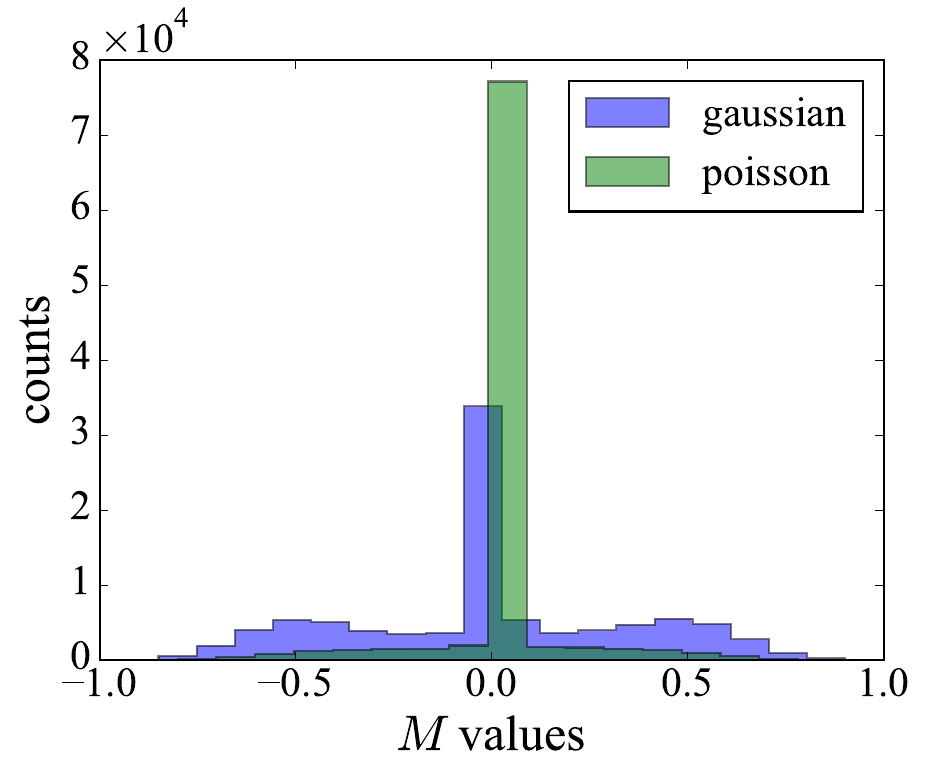}
\\
\hline
$10^8$ & \includegraphics[width=1 \linewidth]{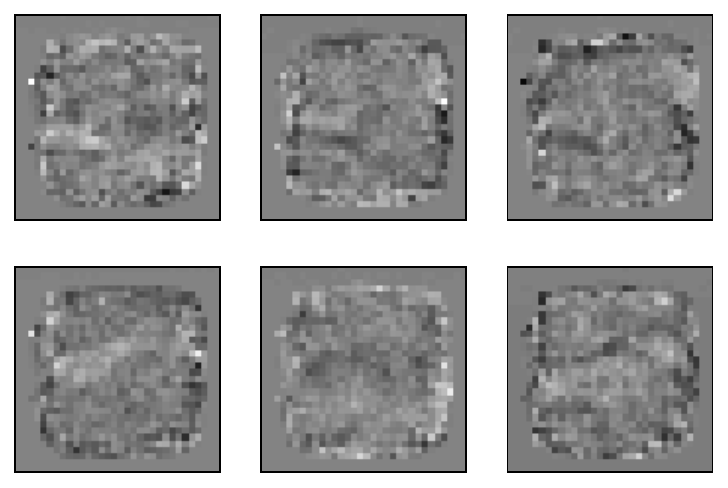} & 
\includegraphics[width=1\linewidth]{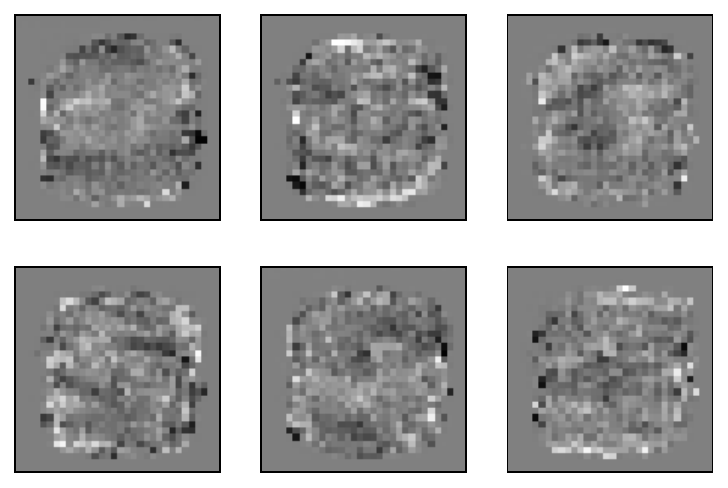} &
\includegraphics[width=1\linewidth]{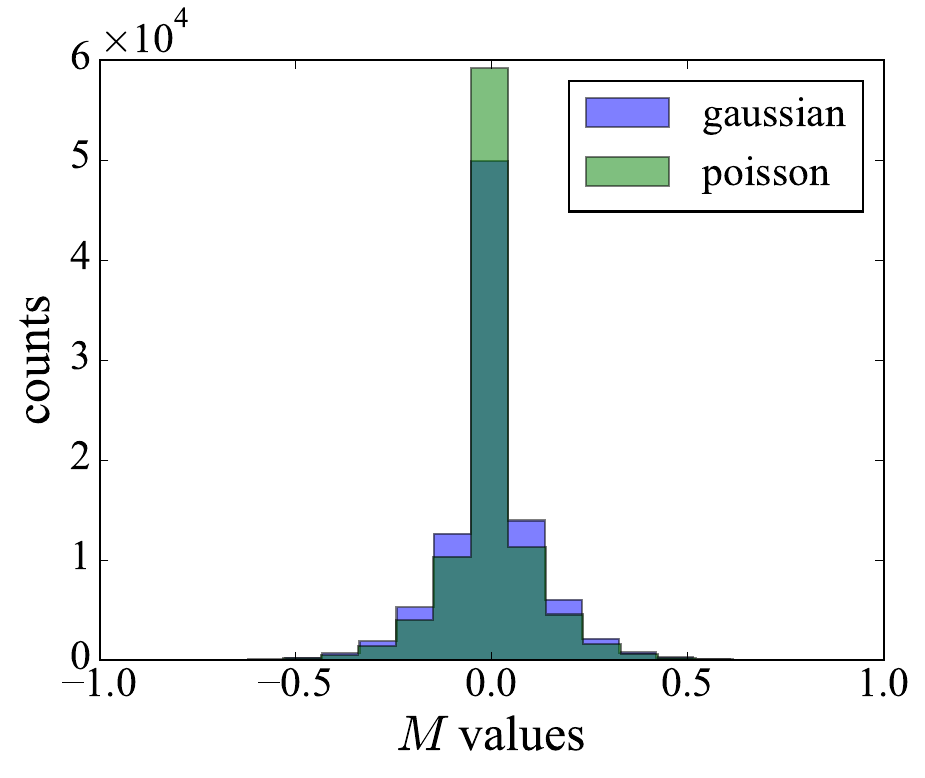}
\\
\end{tabular}
\normalsize
\caption{
    \YLnote{
        Masks optimized by DeepFSI on the MNIST dataset are shown. Each row displays six of the optimized masks under Gaussian and Poisson noise, along with their value distributions for a given number of photons. 
The intensity values range continuously from black to white, with black pixels near $-1$, gray pixels at $0$, and white pixels near $1$.
}
}
\label{fig:MNIST_masks}
\end{figure*}

\YLnote{
\Xpolish{Fig.}{Figure}  
\ref{fig:MNIST_masks} illustrates several DeepFSI-optimized masks for the MNIST dataset. 
%
At low \Xpolish{light levels}{photon budgets}, the mask value distributions for AGN and Poisson noise differ significantly, with Poisson masks focusing on the essential regions and performing repeated measurements to improves the SNR of those informative features, while maintaining lower light throughput compared to Gaussian masks.
However, as the photon count increases, both masks converge to a similar configuration. 
\YLnote{
    This observation is consistent with the conclusions of Raginsky et al. \cite{willet2009CSPoisson}, 
    demonstrating that DeepFSI inherently learns to mitigate the SNR dilution effect characteristic of photon-starved sensing
}
}

\subsection{Error and Robustness Analysis}
\label{ssec:robustness_analysis}

\subsubsection{Compression Robustness}
\begin{figure*}[ht]
    \begin{subfigure}{0.32\linewidth}
        \centering
        \caption{}        
        \includegraphics[width=\linewidth]{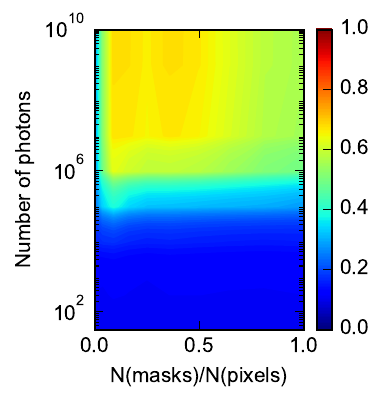}
        \label{fig:truncated_hadamard_2d_classification}
    \end{subfigure}
    \begin{subfigure}{0.32\linewidth}
        \centering
        \caption{}
        \includegraphics[width=\linewidth]{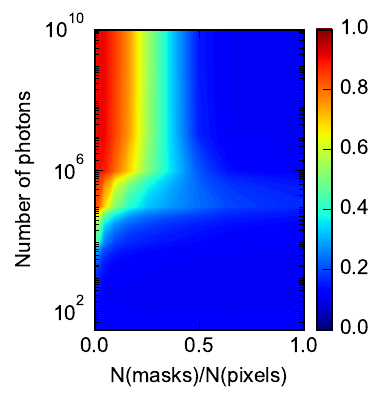}
        \label{fig:hardware_pca_2d_classification}
    \end{subfigure}
    \begin{subfigure}{0.32\linewidth}
        \centering
        \caption{}
        \includegraphics[width=\linewidth]{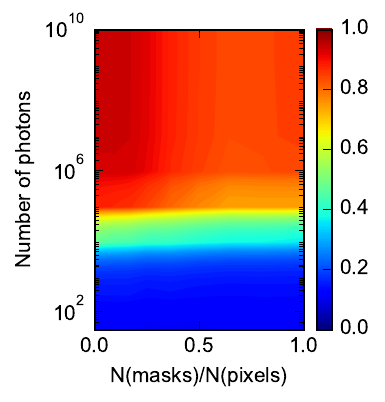}
        \label{fig:onn_2d_classification}
    \end{subfigure}
    \caption{Classification rates on simulated data regarding compression ratios and \Xpolish{light levels}{photon budgets}
    with (\subref{fig:truncated_hadamard_2d_classification}) Truncated Hadamard, (\subref{fig:hardware_pca_2d_classification}) FSI, and (\subref{fig:onn_2d_classification}) DeepFSI.
    The models are trained under Poisson noise using a \Xpolish{light level}{photon budget} of $10^5$ \Xhide{photons} and subsequently tested under different \Xpolish{light levels}{photon budgets}. It is observed that  FSI performance deteriorates when the compression ratio is not properly chosen, whereas the Truncated Hadamard and DeepFSI models exhibit greater robustness in such scenarios.}
    \label{fig:2d_classification_plots}
\end{figure*}

We notice a shortcoming that may affect the practicality of traditional FSI that the optimal number of masks, or the row number in $\B{M}$, is unknown before optimization. 
Furthermore, once $\B{M}$ is optimized, its performance under conditions different from those in the training stage may no longer be optimal.
Therefore, we also design an experiment to test the robustness of TH, FSI and DeepFSI under PN.
For each coding design, we keep using the compression ratios in Section \ref{ssec:simulated_experiments}, but fix the total number of photons at $10^5$ in training stage.
For each compression ratio, we test the optimized coding design with different photon numbers and show the classification rates in Fig. \ref{fig:2d_classification_plots}.

Focusing at the values when number of photon is $10^5$, we can see a tremendous performance degradation in FSI if we choose an improper number of masks, although its 
\YLnote{maximum performance}
is much better than that of non-FSI methods, i.e. the 
\YLnote{\Xpolish{truncated}{Truncated}} \YLnote{Hadamard} 
in this example.
However, while the overall performance is no worse than FSI, DeepFSI presents more robust classification rates when the number of masks is suboptimal.
Given that it is challenging to know the \YLnote{\Xpolish{proper}{appropriate}} number of masks in real-world applications as it depends on the task and training data, DeepFSI has a significant feature of greater tolerance, which greatly improves the practicality compared to FSI.

Focusing at the values with the same compression ratio, we can see the all models' performances are robust when used in less noisy environments where numbers of photons are above $10^5$ in this work, but dramatically drop when used in noisier environments.
We can learn that we should optimize the models with more noise to ensure that their performances are robust in potential situations after setup in practice.

\subsubsection{MLGauss Approximation}



\YLnote{
Since sampling under PN is not directly differentiable, we adopt the MLGauss method during training to enable gradient-based optimization.
To assess the effect of this approximation, we evaluate DeepFSI under both MLGauss and PN models. 
%
}

\begin{figure}[hbt!]
    \centering
    \begin{subfigure}{.49\linewidth}
        \centering
        \caption{MLGauss training + MLGauss testing}
        \includegraphics[width=1\linewidth]{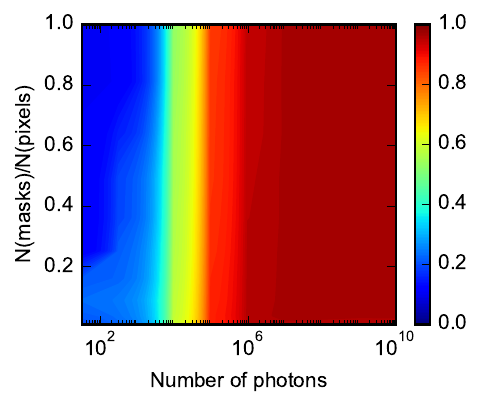}
        \label{fig:mlgauss_train_mlgauss_test}
    \end{subfigure}
    \begin{subfigure}{.49\linewidth}
        \centering
        \caption{MLGauss training + PN testing}
        \includegraphics[width=1\linewidth]{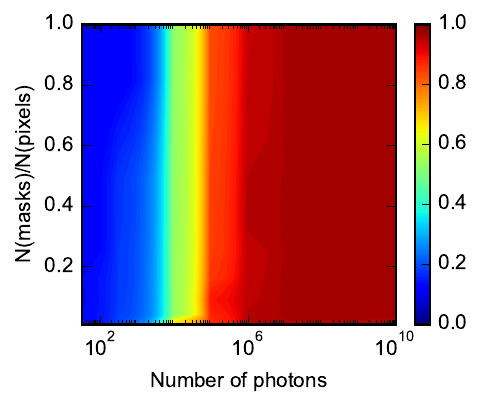}
        \label{fig:mlgauss_train_poisson_test}
    \end{subfigure}
    \caption{
        \YLnote{MLGauss analysis is presented in the contour plots above. Although some degradation is observed when MLGauss is used to train the model for applications under PN, the differences are primarily concentrated in the low light region, where the illumination is insufficient for meaningful classification.}
    }
    \label{fig:mlgauss_analysis}
\end{figure}

\YLnote{
\Xpolish{Fig.}{Figure}   
\ref{fig:mlgauss_analysis} compares performance under matched conditions (MLGauss-training-with-MLGauss-testing) and mismatched conditions (MLGauss-training-with-Poisson-testing).
Noticeable discrepancies between the two testing models mainly arise in the extremely low-photon regime (below 0.1 photons per pixel per measurement), where the overall photon budget is highly limited and classification accuracy is substantially reduced. 
%
Outside this regime, the difference between MLGauss and Poisson testing remains small, suggesting that MLGauss serves as a reasonable and practical approximation to Poisson noise in our setting. 
}

\subsubsection{Quantization}
\YLnote{
We also analyzed the effect of quantization error to evaluate the transfer of the simulated model to practical implementation on a DMD. 
In our system, the masks are quantized to the instrument accuracy of $1/768$, making it important to assess whether this level of quantization affects performance. 
} 

\begin{figure}[hbt!]
    \centering
    \begin{subfigure}{.49\linewidth}
        \centering
        \caption{No quantization under PN}
        \includegraphics[width=1\linewidth]{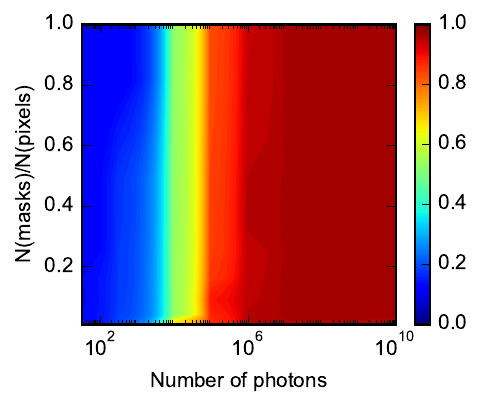}
        \label{fig:quantization_analysis_no_quantization}
    \end{subfigure}
    \begin{subfigure}{.49\linewidth}
        \centering
        \caption{Quantization under PN}
        \includegraphics[width=1\linewidth]{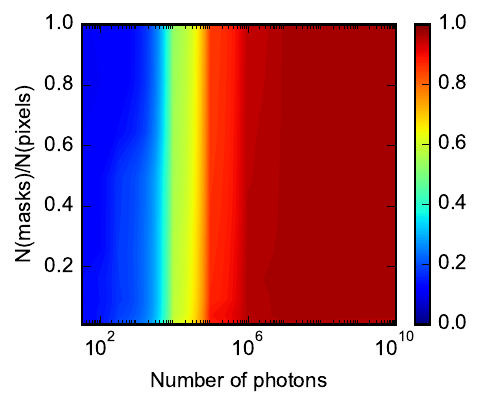}
        \label{fig:quantization_analysis_quantization}
    \end{subfigure}
    \caption{
        \YLnote{Quantization error analysis is performed by comparing the classification rates on DeepFSI with and without quantization, as shown in the contour plots above. The difference between the two modes is not readily apparent.}
    }
    \label{fig:quantization_analysis}
\end{figure}

\YLnote{
\Xpolish{Fig.}{Figure}  
\ref{fig:quantization_analysis} compares DeepFSI performance with and without quantization using the MLGauss-training-with-Poisson-testing mode. 
The results show that quantization has minimal impact on classification accuracy, suggesting that the model is robust to the DMD's discretization limits.
}

\subsection{OViT Simulations}

\begin{figure}[ht]
    \centering    
        \includegraphics[width=.5\linewidth]{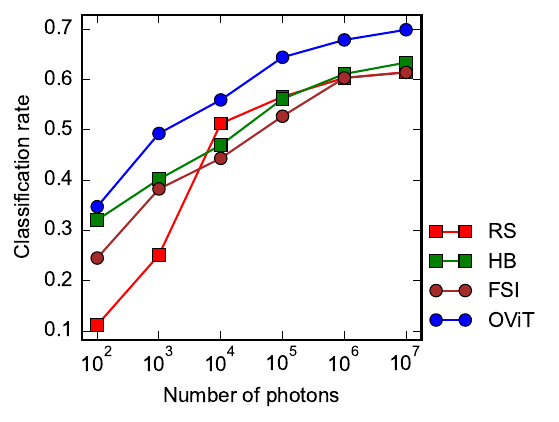} 
    
        \caption{
        Classification rates of different optical coding with ViT on CIFAR-10 under Poisson noise. 
        RS: Raster basis. HB: Hadamard basis. FSI: PCA basis. OViT: Optical-frontend Vision Transformer. 
        The x-axis indicates the total number of photons used in the corresponding simulation.
        }
    \label{fig:OVit_results}    
\end{figure}

\Xpolish{Fig.}{Figure}  
\ref{fig:OVit_results} shows the classification rates on the CIFAR-10 dataset with PN, based on simulation results. 
In this experiment, we assume the best compression rate is unavailable, and all coding designs employ $m=N$ masks where $N$ is the number of pixels in a patch.
When PN is present, OViT outperforms all 
\YLnote{non-supervised}  
coding strategies. 
%
\YLnote{
    This result shows the feasibility of DeepFSI in more complex vision tasks. 
It also verifies its robustness to compression ratio mismatch, as analyzed in Section \ref{ssec:robustness_analysis}, suggesting that DeepFSI is generally more practical than FSI.
}
Notably, our optical scanner module is simply a fully connected layer, much smaller in size compared to the ViT classifier. 
This further underscores that any software cannot ever fully compensate for the performance limitations of suboptimal hardware.
This supports our perspective that once data are measured with suboptimal coding in any vision system, its overall performances on subsequent computer vision tasks are strictly bounded.
All modern computer vision algorithms, unless they also optimize the hardware, are constrained by the upper bound set by the hardware and cannot exceed it, preventing them from fully exploiting their true performance potential.


\begin{figure}[hbt!]
\centering
    \begin{subfigure}{.32\linewidth}
        \centering
        \caption{$10^2$ photons}
        \includegraphics[trim={2cm 1cm 2cm 1cm}, clip, width=\linewidth]{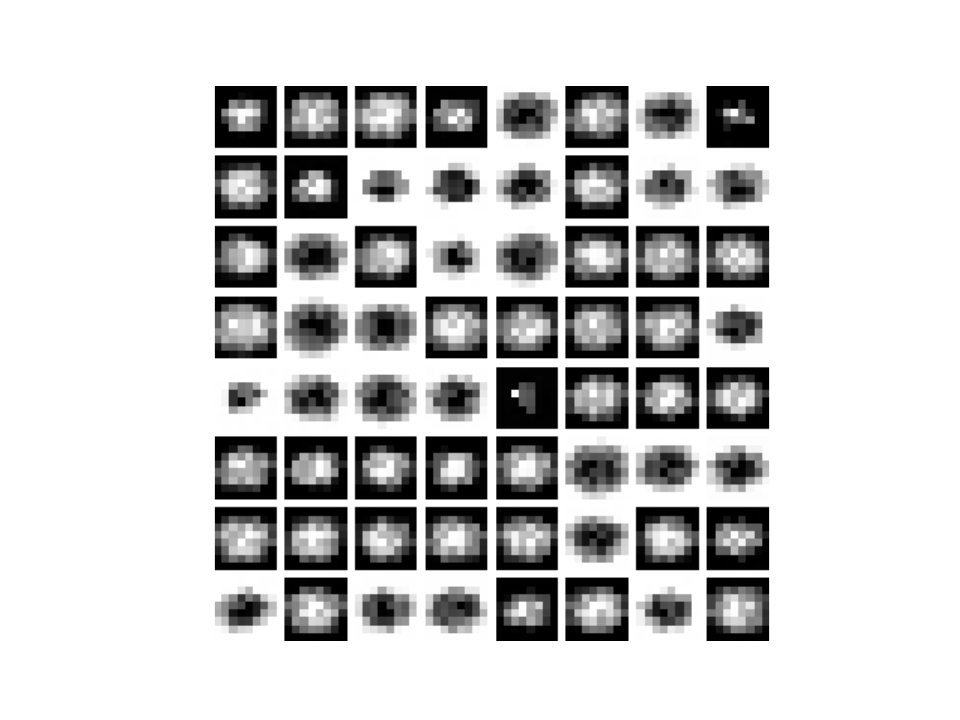} 
    \end{subfigure}
    \begin{subfigure}{.32\linewidth}
        \centering
        \caption{$10^4$ photons}
        \includegraphics[trim={2cm 1cm 2cm 1cm}, clip, width=\linewidth]{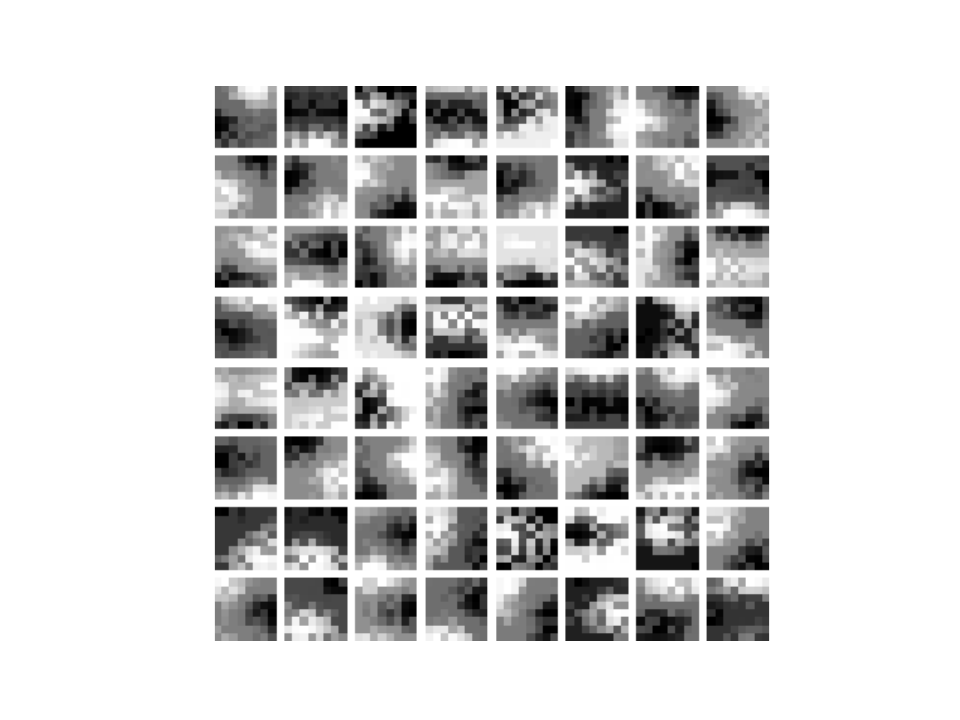} 
    \end{subfigure}
    \begin{subfigure}{.32\linewidth}
        \centering
        \caption{$10^6$ photons}
        \includegraphics[trim={2cm 1cm 2cm 1cm}, clip, width=\linewidth]{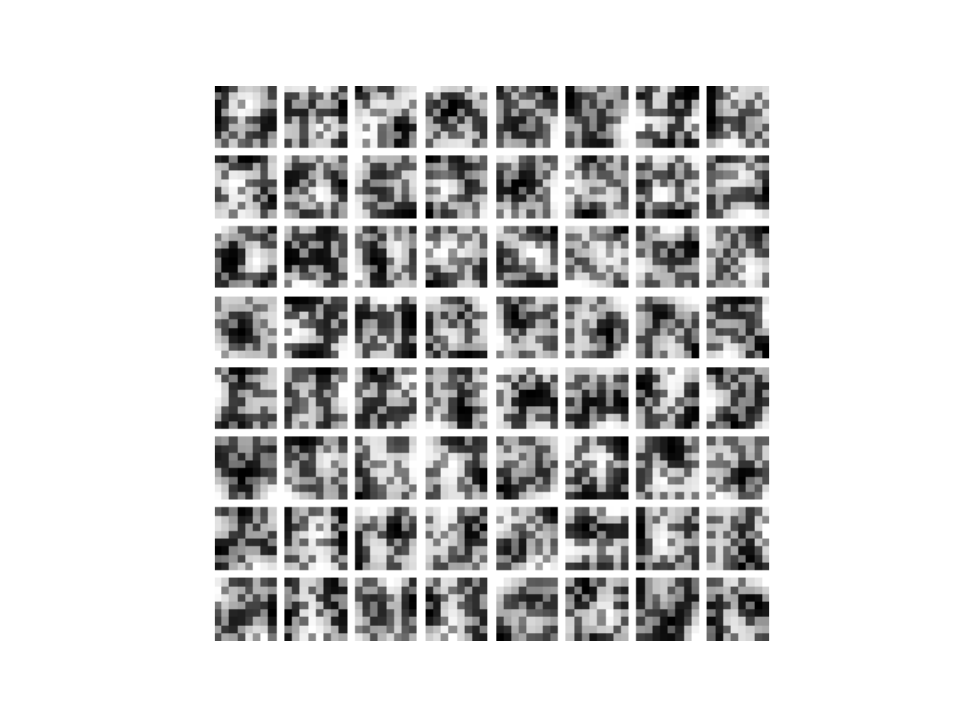} 
    \end{subfigure}
    \caption{ \YLnote{
        Visualization of masks optimized by OViT. 
    Image patches are processed by the 8 by 8 masks sequentially.
    }
    }
    \label{fig:OVit_masks}  
\end{figure}


\YLnote{
\Xpolish{Fig.}{Figure}   
\ref{fig:OVit_masks} shows the OViT-optimized masks with 
different \Xpolish{light levels}{photon budgets}.
The \Xhide{Selective Sensing}{DeepFSI} model always prioritizes the most informative region in signals, and remains frugal in photon counting by reducing the light throughput. 
%
When the photon budget is constrained, DeepFSI refines the quality of selected features through repeated measurements with similar masks, rather than acquiring additional features. 
This behavior is consistent with the pattern previously shown in Fig. \ref{fig:MNIST_masks}.
}

\subsection{Hardware Experiments on MNIST}

\begin{figure}[ht]
    \centering    
        \includegraphics[width=.5\linewidth]{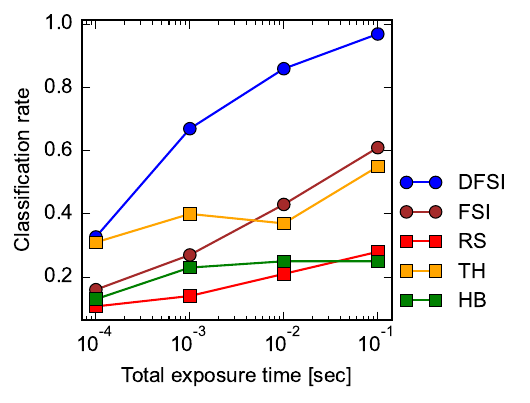} 

        \caption{
            \YLnote{
        Classification rates of different optical coding on MNIST in the hardware experiment. 
        RS: Raster basis. HB: Hadamard basis. TH: \YLnote{\Xpolish{truncated}{Truncated}} Hadamard basis. FSI: PCA basis. DFSI: DeepFSI. 
        The x-axis represents the total exposure time.
            }
        }
    \label{fig:hardware_MNIST_results}    
\end{figure}

\begin{figure}[hbt!]
    \centering    
    \begin{subfigure}{.45\linewidth}
        \centering
        \caption{Exposure time [0.1 sec]}
        \includegraphics[width=1\linewidth]{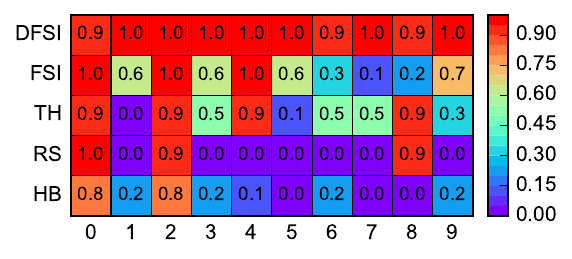}
    \end{subfigure}
    \begin{subfigure}{.45\linewidth}
        \centering
        \caption{Exposure time [0.01 sec]}
        \includegraphics[width=1\linewidth]{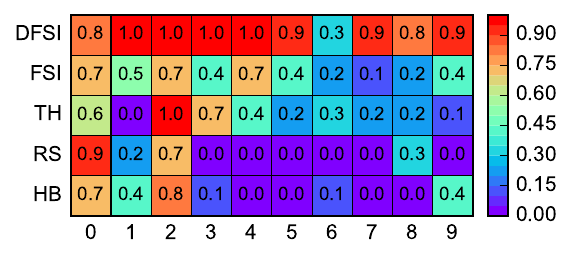}
    \end{subfigure}
    \begin{subfigure}{.45\linewidth}
        \centering
        \caption{Exposure time [0.001 sec]}
        \includegraphics[width=1\linewidth]{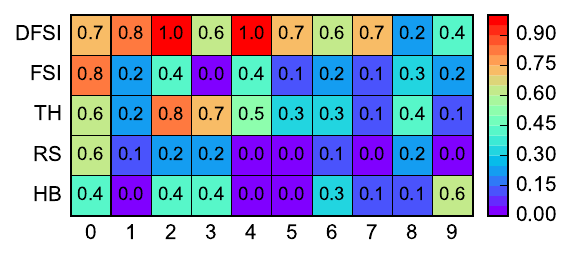}
    \end{subfigure}
    \begin{subfigure}{.45\linewidth}
        \centering
        \caption{Exposure time [0.0001 sec]}
        \includegraphics[width=1\linewidth]{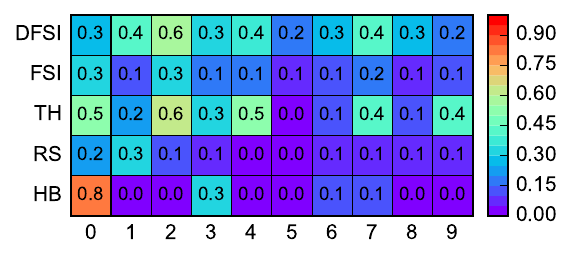}
    \end{subfigure}

    \caption{
        \YLnote{Digit-wise classification rates. Each square shows the classification probability on corresponding digit.}
        }
    \label{fig:digit_specific_classification}
\end{figure}

\Xpolish{Fig.}{Figure}   
\ref{fig:hardware_MNIST_results} compares the classification rates of different coding designs in a hardware experiment.
\YLnote{
    The digit-specific classification performances are illustrated in Fig. \ref{fig:digit_specific_classification}.
}
%
\YLnote{DeepFSI clearly outperforms the other coding designs in all digits when photon budget is sufficient.}
This phenomenon can be explained by the results in Section \ref{ssec:robustness_analysis} that DeepFSI provides better tolerance and performance than the other coding designs when the compression ratio used is not ideal or not available, which is fixed at 0.09 in this experiment.
%
In real-world applications, noise can be more complex than what is considered in this work. 
Factors such as ambient light and dark counts can cause the vision system to deviate from its \YLnote{\Xpolish{supreme}{best performances}} achieved in training conditions, leading to unexpected performance degradation to other coding designs shown in Fig. \ref{fig:hardware_MNIST_results}.
\YLnote{Unfortunately, these factors cannot be fully addressed in the system design, and even achieving the precise number of photons required for coding optimization is rarely possible.}
Therefore, the robustness and tolerance of DeepFSI are of great value in the promotion of task-specific imaging.

\YLnote{
    To strictly evaluate the robustness of our method and avoid pseudoreplication, we performed statistical analysis at the 
    image (cluster) level
    rather than treating individual repetitions as independent samples. Confidence intervals (95\% CI) were calculated using cluster bootstrap resampling ($N=10$ digit classes), and statistical significance was determined using 
    double-sided Wilcoxon signed-rank test
    to account for class-specific difficulty variations in table \ref{table:CI_bootstrap},\ref{table:CI_t_bootstrap}, and \ref{table:Wilcoxon_p_value}.
    %
    The results suggest that DeepFSI achieves competitive and generally stable performance across digit classes and noise levels. 
    Compared to other methods, DeepFSI exhibits smaller performance variations, 
    indicating improved tolerance to mismatches between simulation and experimental conditions. 
    While these findings do not constitute definitive statistical proof due to the sample size, they provide supportive evidence of the \Xpolish{proposed method's}{DeepFSI's} transferability and robustness in practical settings.
%
}

\begin{table}[htbp!]
\tiny 
\centering
\caption{\YLnote{Confidence Interval by Cluster Bootstrap (mean, [lower, higher])}}
\label{table:CI_bootstrap}
\begin{tabular}{r|l|l|l|l|l}  \toprule
Exposure [sec] & DeepFSI                 & FSI                     & Truncated Hadamard      & Raster                  & Hadamard                \\ \hline
0.1                              & $0.970, [0.940, 1.000]$ & $0.610, [0.420, 0.800]$ & $0.550, [0.350, 0.740]$ & $0.280, [0.000, 0.560]$ & $0.250, [0.090, 0.440]$ \\ 
0.01                             & $0.860, [0.720, 0.960]$ & $0.430, [0.300, 0.560]$ & $0.370, [0.200, 0.560]$ & $0.210, [0.030, 0.420]$ & $0.250, [0.080, 0.440]$ \\ 
0.001                            & $0.670, [0.520, 0.810]$ & $0.270, [0.150, 0.410]$ & $0.400, [0.260, 0.540]$ & $0.140, [0.050, 0.260]$ & $0.230, [0.110, 0.360]$ \\ 
0.0001                            & $0.327, [0.258, 0.405]$ & $0.160, [0.117, 0.213]$ & $0.310, [0.190, 0.430]$ & $0.107, [0.057, 0.168]$ & $0.130, [0.010, 0.300]$ \\ 
\hline
\end{tabular}
\end{table}

\begin{table}[htbp!]
\tiny \centering
\caption{\YLnote{Confidence Interval by Studentized Bootstrap (mean, [lower, higher])}}
\label{table:CI_t_bootstrap}
\begin{tabular}{r|l|l|l|l|l}  \toprule
Exposure [sec] & DeepFSI                 & FSI                     & Truncated Hadamard      & Raster                  & Hadamard                \\ \hline
0.1                              & $0.970, [0.939, 0.998]$ & $0.610, [0.337, 0.830]$ & $0.550, [0.295, 0.785]$ & $0.280, [0.018, 0.582]$ & $0.250, [0.088, 0.737]$ \\
0.01                             & $0.860, [0.515, 0.960]$ & $0.430, [0.278, 0.597]$ & $0.370, [0.192, 0.705]$ & $0.210, [0.027, 0.725]$ & $0.250, [0.055, 0.559]$ \\
0.001                            & $0.670, [0.477, 0.823]$ & $0.270, [0.151, 0.509]$ & $0.400, [0.252, 0.595]$ & $0.140, [0.041, 0.358]$ & $0.230, [0.091, 0.402]$ \\ 
0.0001                            & $0.327, [0.253, 0.434]$ & $0.160, [0.115, 0.291]$ & $0.310, [0.156, 0.446]$ & $0.107, [0.054, 0.233]$ & $0.130, [0.012, 0.793]$ \\ 
\hline
\end{tabular}
\end{table}

\begin{table}[htbp!]
\tiny \centering
\caption{\YLnote{Significance Analysis VS DeepFSI (Double-sided Wilcoxon statistic, [p value], rank‑biserial)}}
\label{table:Wilcoxon_p_value}
\begin{tabular}{r|l|l|l|l} \toprule
Exposure [sec] &  FSI                       & Truncated Hadamard        & Raster                    & Hadamard                  \\ \hline
0.1 & $1.000, [\mathbf{0.017}], 0.944$ & $0.000, [\mathbf{0.012}], 1.000$ & $1.500, [\mathbf{0.010}], 0.933$ & $0.000, [\mathbf{0.002}], 1.000$ 
\\
0.01 & $0.000, [\mathbf{0.002}], 1.000$ & $0.000, [\mathbf{0.012}], 1.000$ & $1.000, [\mathbf{0.004}], 0.964$ & $0.000, [\mathbf{0.002}], 1.000$
\\
0.001 & $3.000, [\mathbf{0.010}], 0.891$ & $5.500, [\mathbf{0.027}], 0.800$ & $0.000, [\mathbf{0.008}], 1.000$ & $2.500, [\mathbf{0.010}], 0.909$
\\
0.0001 & $2.000, [\mathbf{0.006}], 0.927$ & $24.000, [0.770], 0.127$ & $0.000, [\mathbf{0.002}], 1.000$ & $9.000, [0.064], 0.673$
\\ 
\hline
\end{tabular}
\end{table}

\subsection{Experiments on Indian Pines}

\begin{figure}[hbt!]
    \centering
    \begin{subfigure}{.49\linewidth}
        \centering
        \caption{Classification under AGN}
        \includegraphics[width=1\linewidth]{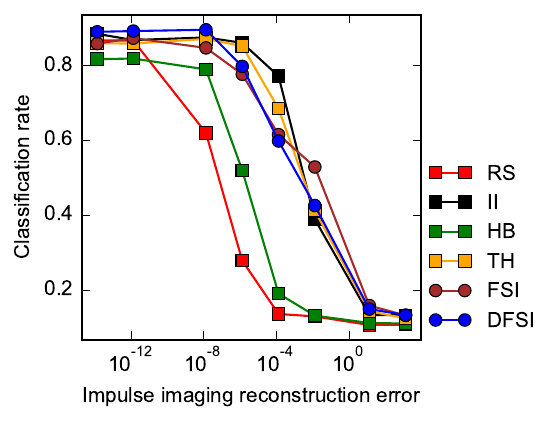}
        \label{fig:gaussian_classification_IndianPines}
    \end{subfigure}
    \begin{subfigure}{.49\linewidth}
        \centering
        \caption{Classification under PN}
        \includegraphics[width=1\linewidth]{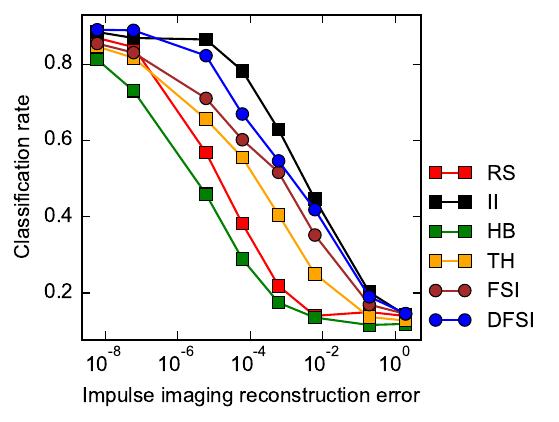}
        \label{fig:poisson_classification_IndianPines}
    \end{subfigure}
    \caption{
    \YLnote{    
    Classification rates on \textbf{Indian Pines}. 
    RS: Raster scan. 
    HB: Hadamard basis. 
    II: Impulse imaging. 
    %
    %
    TH: Low Frequency Truncated Hadamard basis.
    FSI:  PCA basis. 
    DFSI: DeepFSI. 
    %
    %
    Feature-specific methods are marked by $\bigcirc$ from traditional ones marked by $\square$. }
    %
    %
    %
    }
    \label{fig:simulated_classification_IndianPines}
\end{figure}


\YLnote{
    \Xpolish{Fig.}{Figure}   
    \ref{fig:simulated_classification_IndianPines} illustrates the classification performances on Indian Pines. 
    %
    The overall trends are consistent with our observations on MNIST, demonstrating the generalization capability of DeepFSI to hyperspectral data.
    %
    Although DeepFSI does not exhibit clear advantages under AGN assumption, its performance under PN assumption surpasses that of all other methods designed based on the AGN assumption.
    These results indicate that DeepFSI could maintain robust performance in higher-dimensional, more realistic sensing scenarios, highlighting its potential for generalization beyond conventional image classification benchmarks.
}

\section{Discussion and Limitations}
\label{sec:6_discussion}

We propose a deep feature-specific imaging model to complement conventional feature-specific imaging techniques, which is originally designed for additive Gaussian noise, under Poisson noise.
Its performances have been tested with a single-pixel camera compared to other coding designs.
\YLnote{Instead of prioritizing pixel-level reconstruction, DeepFSI focuses on preserving task-relevant information and functional feature fidelity, enabling reliable task performance even when photon levels are insufficient for accurate image reconstruction.}
DeepFSI presents a better performance in classification tasks than that of FSI under Poisson noise and a better practicality and tolerance in real-world implementation as well.
Since Poisson noise is a growing challenge for photon-counting technology in future vision systems design, the proposed model extended the applicability of sensors using such technology.

In addition to some facts already known to many, such as that a higher light throughput may not guarantee a better performance which is shown in Fig. \ref{fig:simulated_classification}, this work also discloses several other interesting phenomena. 
First, using AGN-optimal coding under PN is not recommended as the computed gradient on optical coding layer in these two models are different, indicating different directions during optimization.
Second, future camera or vision system design must incorporate hardware optimization to achieve the best performance.
However, this end-to-end optimization must be performed under the correct noise model.
\YLnote{
    Our results further suggest that in photon-limited regimes, the achievable performance of an FSI system is largely constrained by the efficiency of optical sampling at the front end, rather than solely by the complexity of the back-end model. Optimizing the physical coding layer therefore enables more effective utilization of the available photons, beyond what can be achieved through digital post-processing alone.
}

\YLnote{
    %
    %
    The training complexity of DeepFSI is task specific. 
    For DeepFSI on MNIST, the complete hyperparameter sweep (10 compression ratios, 4 photon budgets, 2 noise models, and 5 trials) required approximately $2.87\times 10^4$ seconds on an RTX 4500 Ada Generation GPU.
    Here, a single run refers to training and evaluating one specific combination of compression ratio, photon budget, noise model, and trial, which required approximately 71 seconds.
    %
    For the hardware experiments, the primary time bottleneck lies in photon timestamp parsing rather than model inference. DeepFSI and FSI processing is relatively efficient, requiring approximately 1 second per run on an AMD Ryzen Threadripper PRO 7965WX (24-core) CPU, including parsing and final recognition. In contrast, Hadamard and \YLnote{\Xpolish{truncated}{Truncated}} Hadamard patterns are significantly more time-consuming due to their higher light throughput, requiring approximately 4 hours for 100 runs.
    Training OViT is more computationally demanding, requiring approximately 2.5 hours per run on a Quadro P5000 GPU. The batch size is selected to maximize GPU memory utilization.
    %
    The influence of training data size on performance has not been systematically evaluated in this work and is left for future investigation.
}

One limitation in this work is that we use the MLGauss noise model whose variance and mean are equal and apply reparameterization to approximate the actual photon noise at the sensor during training.
This approach offers a valid alternative to Poisson noise in gradient estimation when the number of photons is greater than 10 \cite{mitra2014can}, 
%
\YLnote{but it may not provide the same performance gain at extremely low-light conditions, where the actual Poisson noise for testing deviates more from it.}
%
\YLnote{
    In our experimental settings, most methods that use a coding matrix generally operate well once the photon counts are sufficient for classification tasks (with more than one photon from each pixel before coding for each readout), 
    as they combine photons from multiple pixels and exceed this threshold.
    Raster scan, however, may be more affected because it measures only a single pixel at a time. 
    %
    If the photon count falls below the level required for meaningful classification, errors in gradient estimation become noticeable, but in this regime they are a secondary concern since accurate classification is not achievable anyway.
    While we have observed this behavior for DeepFSI as shown in Fig. \ref{fig:mlgauss_analysis}, other coding strategies may show similar trends, though further evaluation would be needed to confirm this.
}
%



\clearpage










\section{Back matter}

\begin{backmatter}
    \bmsection{Funding} 
    U.S. National Science Foundation (1846884); United States Air Force Office for Scientific Research
(FA9550-21-1-0341). This material is based upon work supported by United States Department of Energy and National Nuclear Security Administration under Award Number(s) DE-NA0004196.

    \bmsection{Acknowledgment} \YLnote{The authors would like to thank Sebastian Bauer for insightful discussions that helped identify the limitations of Gaussian assumptions under Poisson noise. We also thank Felipe Gutierrez-Barragan for his valuable contributions in verifying that increased light throughput from coding does not inherently provide advantages under Poisson noise, as well as for assistance with literature review and research planning. We are grateful to Trevor Seets for helpful discussions on research planning and related literature. We thank Ehsan Ahmadi for conducting initial tests on the Indian Pines dataset. Finally, we acknowledge Rebecca Willett for meetings and discussions that helped clarify the fixed photon budget constraint and for verifying the consistency of our preliminary results.}
    
    \bmsection{Disclosures} The authors declare no conflicts of interest.

    \bmsection{Data availability} Data underlying the results presented in this paper are not publicly available at this time but may be obtained from the authors upon reasonable request.

\end{backmatter}

\bibliography{hardware_pca_onn__Main__bibtex.bib}

\end{document}